\documentclass{gGAF2e-mod}

\usepackage{multiequations}

\newcommand*{\be}{\begin{equation}}
\newcommand*{\ee}{\end{equation}}
\newcommand*{\bea}{\begin{eqnarray}}
\newcommand*{\eea}{\end{array}}
\newcommand*{\bal}{\begin{align}}
\newcommand*{\eal}{\end{align}}
\newcommand*{\bme}{\begin{multiequations}}
\newcommand*{\eme}{\end{multiequations}}
\newcommand*{\se}{\singleequation}

\newcommand*{\te}{\tripleequation}

\newcommand\trans{\mbox{tr}}
\newcommand\as{\mbox{a.s.}}

\renewcommand*{\Omega}{\varOmega}
\renewcommand*{\Sigma}{\varSigma}

\newsavebox{\astrutbox}
\sbox{\astrutbox}{\rule[-5pt]{0pt}{20pt}}

\def\squarebox#1{\hbox to #1{\hfill\vbox to #1{\vfill}}}
\newcommand{\reel}{\mathbb R}

\newcommand{\defin}{\stackrel{\scriptscriptstyle\triangle}{=}}

\newcommand{\w}{\boldsymbol{w}}
\newcommand{\bu}{\boldsymbol{u}}
\newcommand{\bphi}{\boldsymbol{\phi}}
\newcommand{\bpsi}{\boldsymbol{\psi}}
\newcommand{\B}{\boldsymbol{B}}
\newcommand{\bsigma}{\boldsymbol{\sigma}}

\newcommand{\xx}{\boldsymbol{x}}

\newcommand{\XX}{\boldsymbol{X}}
\newcommand{\yy}{\boldsymbol{y}}

\newcommand{\YY}{\boldsymbol{Y}}
\newcommand{\nab}{\boldsymbol{\nabla}}
\newcommand{\Delt}{\nabla^2}
\newcommand{\car}{1\!\mathsf{I}}

\newcommand{\transp}{^{\scriptscriptstyle T}}
\newcommand{\tr}{{\scriptscriptstyle T}}

\newcommand{\Exp}{\mathbb{E}}
\newcommand{\Id}{\mathbb{I}}
\newcommand{\f}{\mbs{f}}
\newcommand{\dif}{{\mathrm{d}}}
\newcommand{\mbs}[1]{\ensuremath{\boldsymbol{#1}}}
\newcommand{\mbf}[1]{\ensuremath{\boldsymbol{#1}}}

\begin{document}

\jvol{00} \jnum{00} \jyear{2012} 

\markboth{Fluid flow dynamics  under location uncertainty}{E. M\'emin}

\title{Fluid flow dynamics  under location uncertainty}

\author{E. M\'EMIN$^{\ast}$\thanks{$^\ast$Corresponding author. Email: Etienne.Memin@inria.fr Inria} \\\vspace{6pt} INRIA, Fluminance group, Campus universitaire de Beaulieu, 35042 Rennes Cedex, France\\
\vspace{6pt}\received{Received 11 February 2013; in final form 26 July 2013; first published online ????} }
\maketitle




\begin{abstract}
\noindent We present a derivation of a stochastic model of Navier Stokes equations that relies on a decomposition of the velocity fields into a differentiable drift component and a time uncorrelated uncertainty random term. This type of decomposition is reminiscent in spirit to the classical Reynolds decomposition. However, the random velocity fluctuations considered here are not differentiable with respect to time, and they must be handled through stochastic calculus. The dynamics associated with the differentiable drift component is derived from a stochastic version of the Reynolds transport theorem. It includes in its general form an uncertainty dependent subgrid bulk formula that cannot be immediately related to the usual Boussinesq eddy viscosity assumption constructed from thermal molecular agitation analogy. This formulation, emerging from uncertainties on the fluid parcels location, explains with another viewpoint some subgrid eddy diffusion models currently used in computational fluid dynamics or in geophysical sciences and paves the way for new large-scales flow modeling.
We  finally describe an applications of our formalism to the derivation of  stochastic versions of the Shallow water equations or to the definition of reduced order dynamical systems.\\

\end{abstract}

\section{Introduction}

For several years there has been a deep and growing  interest in defining stochastic models for climate or geophysical sciences \citep{ Majda99, Slingo11}  -- see also on this subject the thematic issue \citep{Palmer08} and references therein or the review \citep{Frederiksen13}. This interest is strongly motivated by the necessity of devising large-scale evolution models that allow taking into account statistical descriptions of processes that cannot be accurately specified - to keep, for instance, an affordable computational time. This includes small-scale physical forcing (cloud convection, boundary layer turbulence, radiative interaction, etc.) and numerical hypothesis and choices operated at the dynamics modeling level, such as a scale coarsening in a given direction, or the selection of a particular functional space to express the solution. The modeling simplifications and the different unresolved processes involved introduce {\em de facto} errors and uncertainties on the constitutive equations of the state variables, which are, in general, so complex that only a probabilistic modeling can be envisaged.

In traditional large-scale modeling, the interaction between the resolved flow component and the unresolved components lies essentially in the constitution of a so-called subgrid stress tensor, which is usually not related to uncertainty or to an error concept. We believe that it is important to extend this notion in order to take into consideration in a more appropriate way the action, on the resolved component, of stochastic processes modeling errors and uncertainties of the state variables' dynamics.

The modeling and the handling along time of such uncertainties are crucial, for instance, for ensemble forecasting issues in meteorology and oceanography, where ensembles of runs are generated through randomization of the dynamics' parameters, accompanied eventually by stochastic forcing that mimics the effect of unresolved physical processes \citep{Slingo11}. Problems related to the underestimation of the ensemble spread and a lack of representativity of the subgrid stress in terms of the model  errors constitute in particular problematic limitations of the ensemble techniques' forecasting skill. Such stochastic evolution models are also needed for data assimilation procedures defined through Monte Carlo implementation of stochastic filters referred to as ensemble filters or particle filters in the literature \citep{Evensen06,Doucet01}. 
In all these situations, a large-scale flow evolution description that includes stochastic forcing terms and an uncertainty related  subgrid stress expression hence appears  necessary.   The root problem of this general issue lies essentially in the construction of adequate  stochastic versions of the Navier-Stokes equations. 
 
The establishment of sound stochastic dynamical models to describe the fluid flow  but also the evolution of the different random terms encoding uncertainties or errors is a difficult issue compounded with an involved mathematical analysis. As initiated by \citep{Bensoussan-Temam-73} and intensively studied by many authors since then  \citep[see for instance the review][and references therein]{Flandoli-08} such a formalization has been mainly considered through the addition of random forcing terms to a standard expression of Navier-Stokes equations. However, this construction is  limited by {\em a priori} assumptions about the noise structure. Furthermore, the question whether this noise should be multiplicative or additive immediately arises.  

Another family of methods initiated by \cite{Kraichnan59} consists to close the large scale representation by neglecting statistical correlations in the Fourier space through the so-called Direct-interaction approximation (DIA). These methods  can be generalized by relying on a Langevin stochastic representation \citep{Kraichnan70, Leith71} to devise advanced stochastic subgrid models for barotropic flows or quasi-geostrophic models where interactions between turbulent eddies and topography are taken into account\citep{Frederiksen99,Frederiksen12,Frederiksen13}. This strategy  based on renormalized perturbation theory comes to randomize the spectral Navier-Sokes representation by replacing the nonlinear interaction terms by appropriate random forcing and damping terms.  

Compared to these works, we wish to explore the problem via a somewhat reverse strategy. Instead of considering a given -- eventually simplified -- dynamics and then to supplement it with random forcing terms to model errors carried by unresolved or unknown processes, we will start from a general Lagrangian stochastic description that incorporates uncertainties on the fluid motion. The sought-after Eulerian dynamics is then deduced from this general stochastic velocity description and standard physical principles or approximations. This construction, reminiscent to the framework proposed by \citep{Mikulevicius04} and initiated by \citep{Brzezniak91}, has the great advantage to let naturally emerge deterministic and stochastic uncertainty terms related to the different errors transported by the evolution model. Deterministic approximations or stochastic implementations of the dynamics can then be considered on solid grounds. 

 
Following this route, we aim in this paper at devising stochastic dynamics for the description of fluid flows under uncertainties. Such uncertainties, modeled through the introduction of random terms, allow taking into account approximations or truncation effects performed during the dynamics constitution steps. This includes for instance the modeling of the unresolved scales interaction in Large-Eddies Simulations (LES) or Reynolds Average Numerical Simulations (RANS). These uncertainties may also encode numerical errors, unknown forcing terms or uncertainties on the initial conditions.  They gather the effects of processes one does not wish to accurately model. However, as they propagate along time and interact with the resolved components their effects must be properly taken into account. 

We will assume throughout this study that the fluid particles displacements can be separated in two components: a smooth differentiable function of time and space and an uncertainty function uncorrelated in time but correlated in space. This latter component is formulated  as a function of Brownian motion and the whole displacement is defined as an Ito diffusion of the form: 
\begin{align}
\label{Ito1}
\dif\XX(\xx,t)= \w(\XX(\xx,t),t)\dif t +\bsigma(\XX(\xx,t),t)\dif\hat \B_t,
\end{align} 
where $\XX:\Omega\times\mathbb{R}^{+}\rightarrow \Omega$ is the fluid flow map, which represents the trajectory followed by a fluid particle starting at point $\XX_{|t=0}(\xx)=\xx$ of the bounded domain $\Omega$. This constitutes a Lagrangian representation of the fluid dynamics and $\dif\XX(\xx,t)$ figures the Lagrangian displacement map of the flow at time $t$.  In expression (\ref{Ito1}), the velocity vector field, $\w$, corresponds to the smooth resolved velocity component of the flow. It is assumed to be a deterministic function (of random arguments) and to have twice differentiable components: $(u,v,w) \in C^2(\Omega,\reel)$. When it does not depend on random arguments this velocity field represents the expectation of the whole random velocity field. This is the situation encountered in the case of  mean field dynamics. The second term is a generalized random field that assembles the unresolved flow component and all the uncertainties we have on the flow. The combination of both velocity fields provides an Eulerian description of the complete velocity fields driving the particles:
\begin{equation}
\label{sto-velocity-map}
\mbs U(\xx,t)  = \w(\xx,t)\dif t + \bsigma(\xx,t) \dif\hat \B_t.
\end{equation}
The whole random field, $\boldsymbol U(\xx,t)$, should be a  solution of the Navier-Stokes equations and is defined here as the combination of a stochastic uncertainty component and a deterministic ``resolved'' component driven respectively by unknown characteristics,  $\bsigma$ and $\w$ that have to be determined or specified. 

This framework is also related to the work of   \citep{Constantin08}. Nevertheless, their study aimed at building a Monte-Carlo Lagrangian representation of the Navier-Stokes equations. It is limited to a constant diffusion tensor $\sigma$ and exhibits additional difficulties to deal with boundary conditions \citep{Constantin11}. In this work, we do not look for a stochastic Lagrangian representation of the deterministic Navier-Stokes equation. We seek instead, a deterministic smooth representation of the drift velocity component corresponding to a solution of a stochastic Navier-Stokes  equation. The goals are thus in some way opposite. Our objective can be interpreted in terms of  Large Eddies Simulation (LES) or Reynolds Average Numerical Simulation (RANS), as we aim at separating a resolved flow component from an unresolved one, respectively defined as a differentiable component and a time uncorrelated uncertainty component.
 From this prospect, the approach proposed here is closer in spirit to RANS techniques than to the LES paradigm as no spatial filtering is applied (see \citep{Lesieur96, Meneveau00, Sagaut04}, for extensive reviews on the subject). Furthermore, let us point out that the uncertainty component lives at all the hydrodynamical scales.  Thus, there is no spatial scale separation principle here but rather a temporal decomposition in terms of a highly oscillating process with no time differentiation property  and a smooth differentiable component. This approach is also close in spirit to the separation in term of a "coherent" component plus noise operated through adaptive wavelet basis \citep{Farge-PRL01,Farge99}. However, contrary to this approach relying on a Galerkin projection with an adaptive scale thresholding, our decomposition makes appear a diffusion tensor  assembling the action of the unresolved uncertainty component on the resolved component.
 
 More precisely, the resolved component of the Navier-Stokes equation we consider includes in its general form an anisotropic diffusion term that emerges due to the presence of the random uncertainty term. By analogy with the Reynolds decomposition and LES, we refer to this tensor as the subgrid stress tensor. However, it is important to outline that its construction differs completely from the Reynold stress definition. As a matter of fact, in our case the unresolved fluctuating component is a non differentiable random process, and stochastic calculus differentiation rules have to be used.  We will see that in the general case the resulting subgrid stress cannot be immediately related to the usual eddy viscosity assumption, formulated in the nineteen century by  \citep{Boussinesq77} from thermal molecular agitation  analogy -- also commonly referred to as {\it Boussinesq assumption} in the litterature --  and that is still intensively used  in the Large Eddies Simulation paradigm since the work of  \citep{Smagorinsky63} and \citep{Lilly66}. 
 
 The simpler form of eddy diffusion will be recovered only when the uncertainty will be confined to homogeneous random fields such as the Kraichnan random field \citep{Kraichnan68}.  For particular forms of the uncertainty random component, this formulation will explain with another viewpoint some subgrid eddy diffusion models proposed  in computational fluid dynamics or in geophysical sciences. Following this route, an appropriate definition of uncertainty models adapted to given situations allows  opening  the way  for new large scales  flow modeling. 
 
 In a very similar way as  in the deterministic case, the representation we propose  will be determined from a stochastic version of the Reynolds transport theorem relying on a specific model of the uncertainty random field.
 
 \section{Construction of the uncertainty random field}
  Before presenting in detail the derivation we consider to built stochastic fluid flow evolution models, we need to specify the random component that will encode the uncertainty associated to the fluid parcels location. This random field, which lives on the bounded domain $\Omega$, relies on a Brownian motion field, refer hereafter to as Brownian avatar (as a smooth idealization of a dense Brownian map) 
defined on $\reel^d$.  This Brownian field is defined in the following section. It is built from a finite dimensional discrete set of standard Brownian variable and tends in law to a continuous limit. This construction will allow us to handle  the spatial derivatives of the uncertainty field with simple calculus. 
 
\subsection{Brownian motion field avatar}
This random field, denoted $\hat \B_t:\reel^d \rightarrow \reel^d$, is built from a finite dimensional discrete set of standard Brownian variables as 
\begin{equation}
\hat \B^n_t(\xx)= \frac{1}{\sqrt n} \sum_{i=1}^n  \B_t(\xx_i)\varphi_\nu (\xx-\xx_i),
\end{equation}
 where $\B_t=\{\B_t(\xx_i),\, i=1,\ldots,n\}$ is a set of independent $d$-dimensional (with $d=2$ or 3) standard  Brownian motions centered on the points of a  discrete grid $S=\{\xx_i,\,i=1,\ldots,n\}\subset \Omega$ and $\varphi$ is a Gaussian function of standard deviation $\nu$. It is immediate to check that  $\hat \B^n_t$ is a zero mean Gaussian process, with uncorrelated increments.
 Its spatial covariance tensor is defined as
\begin{align}
\Exp\bigl[\hat{\B}^n_t(\xx) \hat{\B}^{n\tr}_t(\yy)\bigr]&= \frac{t}{n}\Id_d\sum\limits_{i} \varphi_\nu(\xx-\xx_i)\varphi_\nu(\yy-\xx_i), \end{align}
and, for an infinite number of grid points, tends  to
\begin{equation}
\label{limit-cov}
\mbf Q=\lim_{n\rightarrow \infty}\Exp\bigl[\hat{\B}^n_t(\xx) \hat{\B}^{n\tr}_t(\yy)\bigr] = t\varphi_{\!\sqrt{2}\nu}(\xx-\yy)\Id_d,
\end{equation}  
as
\begin{subequations}
\begin{align}  
\lim_{n\rightarrow \infty}\frac{1}{n}\sum\limits_{i} \varphi_\nu(\xx-\xx_i)\varphi_\nu(\yy-\xx_i) &= \bigl(4\pi \nu^2\bigr)^{-d/2}\exp\Bigl(-\frac{1}{4\nu^2}\|\xx-\yy\|^2\Bigr)\\
&=\varphi_{\!\sqrt{2}\nu}(\xx-\yy). 
\end{align}
\end{subequations}
 It can be checked the operator 
 \begin{equation}
 \mbf Q f(\xx) = \int_{\reel^d} \mbf Q(\xx,\yy) f(\yy) \dif \yy,
 \end{equation}
 is symmetric positive definite. In addition, in  order to define a well defined Gaussian process the trace of this covariance must be bounded for any   orthonormal basis $\{e_k | k\in \mathbb{N}\}$ of the associated function space. It can be checked this covariance operator is not of finite trace in $L^2(\reel^d)$. However it is well defined in a bigger space (that includes in particular the constant functions) such as the space of mean square integrable functions  $L^2((\reel^d),\mu(\xx))$ associated to the measure $\dif\mu(\xx)= \dif \xx/(1+|\xx|^\alpha)$. As a matter of fact, denoting $(,)_\alpha$ the inner product  associated to the space $L^2(\reel^d,\mu(\xx))$, we have  for any orthonormal basis $\{e_k | k\in \mathbb{N}\}$ of  this space
\begin{subequations}
\begin{align}
\trans \; \mbf Q \defin \sum_{i=1}^\infty \left( Q e_i, e_i\right )_\alpha &= td\sum_{i=1}^\infty \int_{\reel^d}  \int_{\reel^d}\varphi_{\sqrt{2}\nu}(\xx-\yy) e_i(\xx)e_i(\yy) \dif\mu(\yy) \dif\mu(\xx)\\
&\leq td\sum_{i=1}^\infty \left(\int_{\reel^d}\varphi_{2\nu} (0) e_i(\yy) \dif\mu(\yy) \right) \left(\int_{\reel^d}\varphi_{2\nu} (0) e_i(\xx) \dif\mu(\xx)  \right)\\
 &\leq td\sum_{i=1}^\infty \bigl(\varphi_{2\nu} (0),e_i\bigr)_\alpha^2\\
 &\leq  td \int_{\reel^d}\varphi^2_{2\nu} (0) \dif\mu(\xx) \leq C, \qquad \mathrm{when}\quad \alpha > d-1.
\end{align}
\end{subequations}
As $\hat{\B}^n_t$ is a Gaussian process, it tends thus in law to a zero mean continuous process in $L^2(\reel^d,\mu(\xx))$ with the same limiting covariance \footnote{Note the covariance operator would be also well defined in a periodic domain $L$. In this case $\trans\, \mbf Q= \|\varphi_\nu\|^2_L$.}. This limiting process will be  denoted in a formal way through a convolution product
\begin{equation}
\B_t\star \varphi_\nu (\xx) = \int_{\reel^d} \B_t(\xx')\varphi_\nu (\xx-\xx')\dif \xx',
\end{equation}
and will be identified to $\hat{\B}_t$ in the following.  Furthermore, it can checked that the covariance of  $\hat \B^n_t$ has a  finite trace. Indeed, for  any orthonormal basis $\{e_k | k\in \mathbb{N}\}$ of  $L^2$, denoting  by $\big (, \big )$ the associated  inner product  and $|f|_2= (f,f)^{1/2}$ its induced norm, we have
\begin{subequations}
\begin{align}
\trans \;  \Exp \bigl[\hat \B^n_t (\xx) \hat \B^{n\tr}_t (\xx)\bigr]&= \sum_{k=1}^\infty \int_{\reel^d}  \int_{\reel^d} e_k\transp(\xx)\Exp \bigl[\hat \B^n_t (\xx) \hat \B^{n\tr}_t (\yy)\bigr]  e_k(\yy) \dif \yy  \dif \xx\\
&= \sum_{k=1}^\infty \Exp   (\hat \B^n_t,  e_k   )^2=\Exp \bigl\|\hat \B_t^n \bigr\|^2_2,
\end{align}
\end{subequations}
 where the  Brownian avatar $L^2$ norm is given by
\begin{align}
\Exp \bigl\|\hat \B^n_t \bigr\|^2_2 &= \frac{td}{n}\sum\limits_{i=1}^n \int_{\reel^d}\varphi^2_\nu(\xx-\xx_i)\dif \xx = td\bigl(4\pi \nu^2\bigr)^{-d/2}.
\label{trace-BA}
\end{align}
The energy of the Brownian avatar hence does not depend on the number of grid points used for its construction but only on the standard deviation of the Gaussian smoothing function. The trace of the limiting covariance is thus provided by (\ref{trace-BA}). At finite time horizon,  when the standard deviation tends to infinity, the energy brought by the uncertainty tends  to zero and the SDE  (\ref{Ito1}) behaves almost like a deterministic evolution law. At the opposite,  in the zero limit of the smoothing function standard deviation, the variance -- or the energy -- becomes unbounded and the Ito stochastic integral can be no more defined.  For non zero standard deviation the covariance operator is of finite trace and the Brownian avatar tends to a well defined $Q$-Wiener process \citep{DaPrato} in $L^2(\reel^d,\mu(\xx))$. Let us point out that such Gaussian correlation is probably the most widely used kernel in Geo-statistics and Machine Learning despite its strong unrealistic smoothing characteristics. In our case this kernel will be used in combination with a diffusion tensor.  
\subsection{White noise avatar}
The analogue of the white noise  on the bounded domain $\Omega$  is similarly defined in a formal way from the generalized function $\dif \hat\B_t= \dif \B_t\star \varphi_{\nu}$ 
and, $\bsigma_t$,  a linear bounded deterministic symmetric operator  of the space of mean square integrable functions with null  condition outside the domain interior $\reel^d / \Omega \cup \upartial\Omega$. We will furthermore assume that this operator is Hilbert-Schmidt (for any orthonormal basis of  $\{e_k | k\in {\mathbb N}\}$ of $L^2(\reel^d,\mu(\xx))$  the operator $\bsigma: L^2(\reel^d,\mu(\xx))\rightarrow L^2(\Omega)$ has a bounded operator norm: $\sum\limits_{k\in {\mathbb N}} \|\bsigma e_k \|^2<\infty$).

Then the uncertainty component reads 
\begin{equation}
\bsigma(\xx,t) \dif \hat\B_t= \int_\Omega\bsigma_t(\xx,\yy) \dif \hat\B_t(\yy) \dif \yy.
\end{equation}
 This operator  $\bsigma_t$ is referred in the following to as the diffusion tensor. 
The limiting covariance tensor of the uncertainty component $\bsigma_t \dif \hat\B_t$, (also called the two points correlation tensor)  reads
\begin{align}
\nonumber
\mbs Q&= \lim_{n\rightarrow \infty } \frac{1}{ n} \dif t\,\delta(t-s)  \sum_{i=0}^n \bsigma(\xx,\bullet,t) \star \varphi_\nu(\xx_i)\bsigma(\bullet,\yy,t)\star \varphi_\nu(\xx_i)\nonumber\\&= \dif t\, \delta(t-s) \bsigma_{\varphi_\nu} (\xx,t) \bsigma_{\varphi_\nu} (\yy,t),
\end{align}
where $ {\bsigma}_\nu(\xx,\yy,t)= {\bsigma}(\xx,\bullet,t) \star \varphi_\nu(\yy)$ is a filtered version of $\bsigma$ along its first or second component since it is symmetric. As the diffusion tensor is Hilbert-Schmidt, we observe immediately through Young's inequality that the covariance is of finite trace. It constitutes thus a well defined covariance operator and the resulting process is a $Q$-Wiener process in $L^2(\Omega)$.

Processes of central importance in stochastic calculus are the quadratic variation and co-variation processes. They correspond to the variance and covariance  of the stochastic increments along time. The quadratic co-variation process denoted as $\langle\XX,\YY\rangle _t$, (respectively the quadratic variation for $\YY=\XX$) is defined as the limit in probability over a partition $\{t_1,\ldots,t_n\}$ of $[0,t]$ with $t_1<t_2<\cdots<t_n$, and a partition spacing $\delta t_i= t_i -t_{i-1}$, noted as $|\delta t|_n= \max\limits_{i} \delta t_i$ and such that $|\delta t|_n\to 0$ when $n\to\infty$:
\begin{equation}
\nonumber
\langle\XX,\YY\rangle _t = \lim^P_{|\delta t|_n\rightarrow 0} \sum_{i=0}^{n-1} \bigl(\XX(t_{i+1}) - \XX(t_i)\bigr) \bigl(\YY(t_{i+1}) - \YY(t_i)\bigr)\transp.
\end{equation}
For Brownian motion these covariations can be easily computed and are given by the following rules $\langle B,B\rangle _t =t$, $\langle B,h\rangle _t=\langle h,B\rangle _t = \langle h,h\rangle _t=0$, where $h$ is a deterministic function and $B$ a scalar Brownian motion. In our case, the quadratic variation of the uncertainty component reads
\begin{align}
\biggl\langle \int_0^t   \bigl(\bsigma (\xx,t) \dif\hat \B_t\bigr)^i,\int_0^t \bigl(\bsigma(\xx,t)\dif \hat \B_t\bigr)^j \biggr\rangle& = \nonumber\\
\lim_{n\rightarrow \infty} \frac{1}{n} \int_0^t  \sum_{\ell=0}^n\sum_k \bsigma^{ik}&(\xx,\bullet,s) \star \varphi_\nu(\xx_\ell)\bsigma^{kj}(\bullet,\xx,s)\star \varphi_\nu(\xx_\ell)\dif s\nonumber\\
&= \int_0^t  \sum_k \sigma^{ik}_{\varphi_\nu} (\xx,s) \sigma^{kj}_{\varphi_\nu} (\xx,s)\dif s\nonumber\\
&\defin   \int_0^t a^{ij}(\xx,s) \dif s  \qquad \as
\label{quadvar}
\end{align}
Its time derivative corresponds to the diagonal of the spatial covariance tensor. We can note that for homogeneous random fields $\bsigma(\xx - \yy)$ these expressions simplify. As a matter of fact, in this case, the random component  
$\bsigma(\xx,t) \dif \hat\B_t= \bsigma(\bullet,t) \star \dif \hat\B_t(\xx)$ has a covariance tensor that is defined as
\begin{equation}
 \mbs Q=\dif t\,\delta(t-s) \bsigma(\bullet,t) \star \bsigma(\bullet,t)\star \varphi_{\sqrt{2}\nu}(\xx-\yy).
\end{equation}
It remains thus homogeneous and its quadratic variation tends to a spatially constant $d\times d$ matrix for an infinite number of grid points: 
\begin{align}
\lim_{n\rightarrow \infty}\frac{1}{n}\sum_{i=0}^n [\bsigma(\bullet,t) \star \varphi_{ \nu}(\xx-\xx_i)]^2 \dif t =& \dif t\int_\Omega \left[\bsigma(\bullet,t) \star \varphi_{\nu}(\xx)\right]^2 \dif \xx\nonumber\\
=&   \mbs \Sigma (t) \dif t   \qquad \as
\label{quad-var-hom}
\end{align}
Besides, we remark that spatial derivatives of the white noise avatar can be defined and manipulated in a simple manner:
\begin{align*}
\upartial_{x_i} (\bsigma(\xx) \dif \hat\B_t) =& \lim_{n\rightarrow \infty}\frac{1}{\sqrt n}\sum_{j=0}^n \upartial_{x_i} \bsigma_{\varphi_\nu} (\xx,\xx_j) \dif\B_t(\xx_j)\\ =& \upartial_{x_i} \bsigma_{\varphi_\nu}(\xx)\dif \hat\B_t .
\end{align*}
As a direct consequence, we note that a divergence free random component is thus necessarily associated to a divergence free diffusion tensor. The derivative operator of the identity diffusion tensor  $\bsigma(\xx,\yy)=\delta(\xx-\yy)$ is also still  well defined for a nonzero standard deviation and  acts solely on the white noise avatar Gaussian smoothing function. 

\subsection{Kraichnan turbulent model}
Toys models pioneered by  \citep{Kraichnan68}  and intensively explored for the theoretical analysis of passive scalar turbulence \citep{Gawedzki95,Kraichnan68,Majda-Kramer} can  be easily specified with such a model. The Kraichnan random field model
\begin{equation}
\dif{\boldsymbol{\xi}^\zeta_t}  = {\cal P}  \star \psi^{\gamma}_{\kappa} \star f^\zeta \star \dif\B_t,
\label{Kxi}
\end{equation}
is formally defined from the divergence free projector ${\cal P}$, a power function $f^\zeta (\xx) = C_\zeta \|\xx\|^{\zeta/2}$ for $0<\zeta<2$ and a band-pass  cut-off function, $\psi^{\gamma}_{\kappa}$, on the so-called inertial range that lies at high Reynolds number between the short dissipative scale $\ell_{\scriptscriptstyle  D} = 1/\kappa$ and the large integral scale   $L= 1/\gamma$ at which the forcing takes place. The constant $C_\zeta$ has a dimension of $\ell^{1-({\xi}/{2})}/t$, which follows from dimensional analysis (the energy transfer rate $\epsilon=\Exp ({u^2}/{t})\sim {\ell^2}/{t^3}$ and  $\epsilon \sim C_\zeta^2 {\ell^\xi}/{t}$). The spectral correlation tensor of $\dif\boldsymbol{\xi}^\zeta_t$ is defined as 
\[
 \widehat{\mbf Q}(\mbf{k})_{ij}=  |\mbf{k}|^{-\zeta-d}  \biggl(\delta_{ij} - \frac{k_ik_j}{|\mbf{k}|^2}\biggr)\bigl(\,\widehat{\psi^{\gamma}_{\kappa}}\,\bigr)^2,
\]
and the variance (or time derivative of the quadratic variation process) of this homogeneous random field reads 
\begin{equation}
\dif\bigl\langle{\boldsymbol{\xi}^\zeta_t}(\xx),{\boldsymbol{\xi}^\zeta_t}(\xx)\bigr\rangle_{ij}  = \dif t\, \delta_{ij} \int_\Omega \bigl[{\cal P} \star \psi^{\gamma}_{\kappa} \star   f^\zeta (\xx)\bigr]^2 \dif \xx \qquad\as
\end{equation}
which is constant and given for a band-pass filter, $\widehat{\psi^{\gamma}_{\kappa}}$, defined as a box filter as
\begin{align}
\frac{\dif t \,C_\zeta}{(2\pi)^d} \,\frac{d-1}{d} \,\frac{2\pi^{d/2}}{\Gamma({d}/{2})} \,\zeta^{-1}(L^\zeta - \ell^\zeta_{\scriptscriptstyle D} )\,\delta_{ij}.
\label{quad-var-K}
\end{align}
The square root of this quantity represents the mean absolute value of the velocity field components. It blows up when the integral scale tends to infinity and exhibits a strong dependency on the largest length scale. This injection scale dominates the turbulent energy and corresponds to the distance beyond which velocities are uncorrelated.

We will see that such random fields, which are both homogeneous in space and uncorrelated in time,  lead to a simple form of the dynamics drift term involving an intuitive eddy viscosity term defined from the uncertainty random field variance. However more general random fields will let rise a formulation of the Navier-Stokes equations in which emerges an anisotropic diffusion term that cannot be immediately related to the usual eddy viscosity assumption first formulated in the nineteen century by  \citep{Boussinesq77}\footnote{referred sometimes for that reason as the Boussinesq assumption.} and that is still predominantly used  in the Large Eddies Simulation paradigm.

\section{Stochastic Reynolds transport theorem}
 In a similar way as in the deterministic case, our derivation  relies essentially on a stochastic version of the Reynolds transport theorem.  The rate of change of a scalar function $q$ within a material fluid volume ${\cal V}(t)$ transported by (\ref{Ito1}) is
\begin{multline}
\dif\!\int_{{\cal V}(t)} \!\!q(\xx,t) \dif \xx=  \int_{{\cal V}(t)}\!\!\biggl\{\dif_t q +\biggl[ {\nab} {\bm\cdot}(q\w) +
\tfrac{1}{2} \| \nab{\bm\cdot}\, \bsigma \|^2  q  - \sum_{i,j} \frac{1}{2} \frac{\upartial^2 }{\upartial x_i \upartial x_j} (a_{ij}q){|_{\nab{\bm\cdot}\bsigma=0}}\biggr] \dif t \biggr. \\\biggl.+ \nab{\bm\cdot}(q \bsigma \dif\hat \B_t)\biggr\}\dif \xx. 
\end{multline}
 In this expression the forth right-hand term must be computed considering the diffusion tensor is divergence free, and the tensor $\mbs a(\xx)$ denotes the uncertainty variance defined in (\ref{quadvar}) and $\dif_t\phi$ denotes the function increment at a fixed point for an infinitesimal time step.

To give some insights on how this expression is obtained, let us consider a sufficiently spatially  regular enough scalar function $\phi$ of compact support, transported by the flow and that vanishes outside ${\cal V}(t)$ and on  $\upartial{\cal V}(t)$. As this function is assumed to be transported by the stochastic flow (\ref{Ito1}), it constitutes a stochastic process defined from its initial time value  $g$: 
\[
\phi(\XX_t,t)= g(\xx_0),
\]
where both functions have bounded spatial gradients. 
The initial function $g:\Omega\rightarrow \reel$ vanishes outside the initial volume ${\cal V}(t_0)$ and on  its boundary.  
Let us stress the fact that  function $\phi$ cannot be a deterministic function. As a matter of fact, if it was the case, its differential would be given by a standard Ito formula
\[
\dif\phi(t,\XX_t)= \biggl(\upartial_t\phi +{\nab}\phi\, {\bm\cdot}\, \w + \frac{1}{2} \sum_{i,j} \dif\bigl\langle X^{i}_t,X^{j}_t\bigr\rangle \frac{\upartial^2 \phi }{\upartial x_i \upartial x_j}\biggr) \dif t+ {\nab}\phi \,{\bm\cdot}\, \bsigma \dif \hat\B_t.
\]
This quantity cancels as $\phi$ is transported by the flow. A separation between the slow deterministic terms and the fast Brownian term would yield a very specific uncertainty lying on the  function iso-values  surfaces  ($\nab\phi\transp\bsigma=0$) or a null uncertainty, which is not the general goal followed here. 

As $\phi$ is a random function, the differential of  $\phi(\XX_t,t)$  involves the composition of two stochastic processes. Its evaluation requires the use of a generalized Ito formula  usually referred in the literature to as the Ito-Wentzell formula \citep[see theorem 3.3.1,][]{Kunita}.
This extended Ito formula incorporates in the same way as the classical Ito formula\footnote{relevant only to express the differential of a deterministic function of a stochastic process} quadratic variation terms related to process $\XX_t$ but also co-variation terms between $\XX_t$ and the gradient of the random function $\phi_t$. Its expression is in our case given by
\begin{align}
\dif\phi(t,\XX_t)= &\,\dif_t\phi +{\nab}\phi \,{\bm\cdot}\, \dif\XX_t +\sum_i \dif\Bigl\langle\frac{\upartial \phi}{\upartial x_i}, \bigl(\bsigma \hat{\B}_t\bigr)^{\!i}\Bigr\rangle \dif t \nonumber+ \frac{1}{2} \sum_{i,j} \dif\bigl\langle X^{i}_t,X^{j}_t\bigr\rangle \frac{\upartial^2 \phi }{\upartial x_i \upartial x_j} \dif t \\= &\,0,
\label{dxi=0}
\end{align}
 It can be immediately checked that for a deterministic function or for uncertainties directed along the  function iso-values, the standard Ito formula is recovered. Otherwise, for a fixed grid point, function $\phi(\xx,t)$ is hence  solution of a stochastic differential equation of the form
\begin{equation}
\dif_t\phi(\xx,t) = v(\mbs x,t)\dif t + \mbs f(\mbs x,t)\,{\bm\cdot}\, \dif\hat \B_t.
\label{dxi_t}
\end{equation}
The quadratic variation term involved in (\ref{dxi=0}) is given through (\ref{quadvar}) as
\begin{equation}
\dif\bigl\langle X^{i}_t,X^{j}_t\bigr\rangle  = a^{ij}(\xx,t),
\end{equation}  and the covariation term reads
\begin{align}
\dif\Bigl\langle\frac{\upartial \phi_t}{\upartial x_i}, \bigl(\bsigma \hat{\B}_t\bigr)^{\!i}\Bigr\rangle &=\lim_{n\rightarrow \infty}\frac{1}{n}\sum_{j}\sum_{\ell=0}^n \Bigl(\frac{\upartial f^j}{\upartial x_i} (\xx,\bullet,t) \star\varphi_{\nu} (\xx_\ell)\Bigl)\bsigma^{ij} (\xx,\bullet,t)\star \varphi_{\nu}(\xx_\ell),\nonumber \\
&= \sum_{j} \frac{\upartial f^j}{\upartial x_i} \star\varphi_{\nu}(\xx,t)   \bsigma^{ij}_{\varphi_{\nu}} (\xx,t).
\end{align}
From these expressions and identifying first the Brownian term and next the deterministic terms  of  equations (\ref{dxi=0}) and (\ref{dxi_t}), we infer 
\begin{align}
\mbs f(\xx,\yy,t)\transp & = - {\nab}\phi(\xx,t) \transp \bsigma(\xx,\yy,t),\nonumber\\
v(\xx,t) &= - {\nab}\phi\,{\bm\cdot}\, \w + \sum_{i,j}\frac{1}{2} a_{ij}\frac{\upartial^2 \phi}{\upartial x_i \upartial x_j} + \nab \phi\,{\bm\cdot}\, \frac{\upartial \bsigma_{\nu}^{\bullet j}} {\upartial x_i} \bsigma^{ij}_{\nu}, \nonumber
\end{align}
and finally get
\begin{align}
\label{SMD}
\dif_t\phi   &=  {\cal L} \phi  \dif t  -  {\nab}\phi\,{\bm\cdot}\,\bsigma \dif\hat\B_t, \\
{\cal L} \phi &= -{\nab}\phi\,{\bm\cdot}\, \w + \sum_{i,j}\frac{1}{2} a_{ij}\frac{\upartial^2 \phi}{\upartial x_i \upartial x_j} + \nab\phi\,{\bm\cdot} \,\frac{\upartial \bsigma^{j\bullet}_{\varphi_{\nu}}}{\upartial x_i} \bsigma^{ij}_{\varphi_{\nu}} \nonumber.
\end{align}
This differential at a fixed point, $\xx$, defines the equivalent in the deterministic case of the material derivative of a function transported by the flow. For interested readers, several conditions under which  linear equations with a multiplicative noise exhibits a unique weak, mild or strong solution are studied in \citep{DaPrato}. In our case, only existence of a weak solution is needed as we will proceed to a formal  integration by part.  
The differential of the integral over a material volume of the product $q\phi$ is given by
\begin{align}
\dif\int_{{\cal V}(t)} \!\!\!\!q\phi(\XX_t,t)\dif \xx&= \dif\int_{\Omega} \!\!q\phi \dif \xx,\nonumber\\
&= \int_{\Omega} \!\bigl(\dif_t q \phi + q \dif_t\phi  +\dif_t\langle q,\phi\rangle \bigr) \dif \xx, \nonumber
\end{align}
where the first line comes from $\phi(t,\xx)= 0 \mbox{ if }\xx\in\Omega\backslash {\cal V}(t)$
and the second one from the integration by part formula of two  Ito processes. 
Hence, from (\ref{SMD}) this differential is 
\begin{equation*}
 \int_{\Omega} \bigl(\dif_t q \phi + q {\cal L} \phi + \nab \phi\, {\bm\cdot}\, \mbf a \nab q \bigr)\dif t\, \dif \xx- \int_{\Omega} q {\nab}\phi\,{\bm\cdot} \, \bsigma \dif\hat \B_t.
\end{equation*}
Introducing ${\cal L}^\ast$ the (formal) adjoint  of the operator ${\cal L}$ in the space $L^2(\Omega)$ with Dirichlet boundary conditions, this expression can be written as
\begin{equation}
\int_{\Omega} \!\!\left(\bigl( \dif_t q + {\cal L}^\ast q -\nab{\bm\cdot} \,(\mbf a \nab q )\bigr)\dif t + \!\nab{\bm\cdot}\bigl(q \bsigma \dif\hat \B_t\bigr) \right)\phi \dif \xx.\nonumber
\end{equation}
With the complete expression of ${\cal L}^\ast$ and  if $\phi(\xx,t) \rightarrow \car_{{\cal V}(t) /\upartial{\cal V}(t)}$, where $\car$ stands for the characteristic  function, we get the extended form of this differential:
\begin{multline}
\int_{{\cal V}(t)}\! \biggl[\dif_t q +\biggl( {\nab} {\bm\cdot}\,(q\w) - \sum_{i,j} \nab{\bm\cdot}\,\Bigl(q \bsigma_{\nu}^{ij} \frac{\upartial \bsigma_{\nu}^{j\bullet}}{\upartial x_i}\Bigr)\biggr.\biggr.\\ \biggl.\biggl.+ \frac{1}{2} \sum_{ij}\frac{\upartial^2}{\upartial x_i\upartial x_j}(a_{ij}q)  
 -\sum_{ij}\upartial_{x_i} a^{ij}\upartial_{x_j}q - \sum_{ij} a^{ij}\upartial^2_{x_i x_j} q\biggr) \dif t + \nab{\bm\cdot}\bigl(q \bsigma \dif\hat \B_t\bigr)\biggr]\dif \xx,\nonumber
\end{multline}
which can be further simplified  as
\begin{multline}
\int_{{\cal V}(t)}\!\biggl[\dif_t q + {\nab} {\bm\cdot}\,(q\w) \dif t - \frac{1}{2} \sum_{ijk} \Bigl(\sigma_{\nu}^{ik}\sigma_{\nu}^{kj}\frac{\upartial^2 q}{\upartial x_i\upartial x_j} + \upartial_{x_j}\sigma_{\nu}^{ik}\upartial_{x_i}\sigma_{\nu}^{kj}q
\Bigr.\biggr.\\\biggl.\Bigl.+2\sigma_{\nu}^{ik}\upartial_{x_i}\sigma_{\nu}^{kj} \upartial_{x_j}q -\upartial_{x_i}\sigma_{\nu}^{ik}\upartial_{x_j}\sigma_{\nu}^{kj}q\Bigr)\dif t  + \nab{\bm\cdot}\bigl(q \bsigma \dif\hat \B_t\bigr)\biggr] \dif \xx.\nonumber
\end{multline} 
At last this expression may be more compactly written as
\begin{align}
\int_{{\cal V}(t)}  \!\biggl[\dif_t q + {\nab} {\bm\cdot}\,(q\w) \dif t - \frac{1}{2} \sum_{ij} \frac{\upartial^2 }{\upartial x_i \upartial x_j} (a_{ij}q){|_{\nab{\bm\cdot}\, \bsigma=0}} \dif t +\tfrac{1}{2} \| \nab{\bm\cdot}\, \bsigma \|^2  q \dif t   + \nab{\bm\cdot}\bigl(q \bsigma \dif\hat \B_t\bigr)\biggr]\dif \xx,\nonumber
\end{align} 
where the third term must be computed by considering  a divergence free diffusion tensor. Let us note it is now straightforward to get the expression of the transport theorem for a  divergence free   turbulent component:
\begin{equation}
\label{div-free-transport}
\begin{split}
&\dif\int_{{\cal V}(t)} q(\xx,t) \dif \xx= \int_{{\cal V}(t)} \biggl[\dif_t q +\biggl(\! {\nab}{\bm\cdot}\,(q\w) -  \frac{1}{2}\sum_{i,j} \upartial^2_{ij} (a_{ij}q)\!\biggr) \dif t + \nab q \,{\bm\cdot}\,\bsigma \dif\hat \B_t\biggr] \dif \xx.
\end{split}
\end{equation}

\section{Mass conservation}
This  expression of volumetric rate of change relation allows us stating
the mass conservation principle under fluid particles location uncertainty. Applying the previous transport theorem to the fluid density $\rho(\xx,t)$, we get a general  expression of the mass variation:
\begin{multline}
\label{mass-conservation}
\int_{{\cal V}(t)}\!\biggl[\dif_t \rho + \biggl(\!{\nab} {\bm\cdot}\,(\rho \w )  -  \tfrac{1}{2}\sum_{i,j}\frac{\upartial^2}{\upartial x_i\upartial x_j}(a_{ij}\rho){|_{\nab{\bm\cdot}\, \bsigma=0}}+ \frac{1}{2}\|\nab{\bm\cdot}\, \bsigma \|^2  \rho \!\biggr) \dif t  + \nab{\bm\cdot}\bigl(\rho \bsigma \dif\hat \B_t\bigr)\biggr]\dif \xx . 
\end{multline}
A mass conservation constraint on the transported volume implies canceling this expression.  Besides, as the volume is  arbitrary this  provides the following  constraint 
\begin{equation}
\label{mass-conservation-1}
\dif_t \rho + {\nab} {\bm\cdot}\, (\rho \w ) \dif t\,= \,\frac{1}{2}\biggl(\sum_{i,j}\frac{\upartial^2}{\upartial x_i\upartial x_j}(a_{ij}\rho){|_{\nab{\bm\cdot}\, \bsigma=0}}-\tfrac{1}{2}\| \nab{\bm\cdot}\, \bsigma \|^2  \rho\biggr) \dif t  - \nab{\bm\cdot}\bigl(\rho \bsigma \dif \hat \B_t\bigr).
\end{equation}
\subsection{Incompressible fluids}
For an incompressible fluid with constant density, canceling separately the slow deterministic terms and the highly oscillating stochastic terms of the mass conservation constraint (\ref{mass-conservation-1}), we obtain
\bme
\be
%
{\nab} {\bm\cdot}\,(\bsigma \dif \hat\B_t) =0,  \qquad\qquad
{\nab}{\bm\cdot}\,\w = \frac{1}{2}\sum_{i,j}\frac{\upartial^2 a_{ij}}{\upartial x_i\upartial x_j}.
\ee
\eme
In the very same way as in any large scale flow decompositions,  we aim here at recovering a physically plausible volume preserving drift component. Imposing to the drift component to be divergence free,  this system boils down  hence to an intuitive system of incompressibility constraints
\bme
\be
{\nab} {\bm\cdot}\,\bigl(\bsigma \dif \hat\B_t\bigr) =0, \qquad\qquad {\nab} {\bm\cdot}\,\w =0,
\ee
\eme
coupled with a less intuitive  additional constraint on the quadratic variation tensor: 
\begin{equation}
\nab{\bm\cdot}\,(\nab{\bm\cdot}\,\mbf a) = 0.
\label{quad-var-constraint}
\end{equation}
For the Kraichnan model (or for any divergence-free homogeneous random fields) this last constraint is naturally satisfied, as its quadratic variation is constant and the system comes down to the classical incompressibility constraint. In the same way, any homogeneous divergent-free random fields associated to  a volume preserving drift leads to mass conservation.
\subsection{Isochoric flows and isoneutral uncertainty}
For divergence-free volume preserving flows (refered as isochoric flow) with varying density  we get  a mass conservation constraint  of the form
\begin{align}
\dif_t \rho + {\nab}\rho\,{\bm\cdot}\,\w \,\dif t  - \frac{1}{2}\sum_{i,j}  \frac{\upartial^2  }{\upartial x_i x_j}  (\rho  a^{ij})\dif t= \nab\rho\,{\bm\cdot}\, \bsigma \,\dif\hat \B_t.
\end{align}
An interesting property emerges if the uncertainty has a much larger  scale  along the density tangent plane than in the density gradient direction.This situation occurs in particular in oceanography or in meteorology where the fluid is stratified and the horizontal scale much larger than the vertical scale. In a such  case, it is essential that the large-scale numerical simulations  preserve the natural fluid stratification and consequently to  define subgrid models with controlled diffusion along the iso-density surfaces.  This behavior can be easily set up by defining the diffusion tensor as  a  diagonal projection operator  so  that the uncertainty $\bsigma \dif \hat\B_t$ lies on the isodensity surfaces: 
\begin{equation}
\bsigma^{ij} = \delta^{ij} -  \frac{\upartial_{x_i}\rho(\xx) \upartial_{x_j}\rho(\yy)}{\|\nabla\rho\|^2  } \delta(\xx-\yy).
\end{equation}
 Together with  the  small slope assumption \cite{Gent90} ($\sqrt{(\upartial_x \rho)^2 \!\!+\!\! (\upartial_y \rho)^2 }\!\!\ll\upartial_z \rho $) the diffusion tensor and the quadratic variation can then be written as a matrix function
\begin{equation}
\mbf a(\xx)= \begin{pmatrix} 1 & 0 &\alpha_x(\xx)\\[0.3em] 0 &1 &\alpha_y(\xx)\\[0.3em]\,\alpha_x(\xx)\quad& \quad\alpha_y(\xx)\quad& \quad|\mbs\alpha(\xx)|^2 \end{pmatrix},
\end{equation}
 where we introduced the neutral slope vector $\mbf \alpha = (\alpha_x,\alpha_y,0)= - (\upartial_x \rho /\upartial_z \rho, \upartial_x \rho /\upartial_z \rho, 0) $.  Owing to the divergence free condition of the diffusion tensor,  this slope vector is necessarily constant along the depth axis $\upartial_z \alpha_x = \upartial_z \alpha_y = 0$ and divergence free $\nab\!{\bm\cdot} \mbf \alpha =0$. This yields for the mass preserving equation a diffusion  along the density tangent plane.  The density evolves as the deterministic model: 
\begin{equation}
\label{isoneutral-mass-conservation}
\frac{\upartial \rho}{\upartial t} + \nab\rho\,{\bm\cdot}\, \w  = \frac{1}{2} \sum_{ij}\upartial_{x_i} (a_{ij} \upartial_{x_j}\rho),
\end{equation}
where we considered the divergence free constraint ${\bm\nabla} {\bm\cdot} \,\mbf a =0$. This type of anisotropic diffusion corresponds exactly to the so-called  isoneutral diffusion currently used to model unresolved mesoscale eddies in coarse scale resolution of ocean dynamics simulations \citep{Gent90}. As mentioned previously, the divergence free uncertainty induces a slope vector that is also divergence free and independent of the depth. This enforces further the density surfaces to be quasi-harmonic (for smooth depth density variation) in the horizontal planes and provides a strong stratification of the density organized as a pile of pancakes with no maxima excepted on the domain horizontal boundaries. To our knowledge those additional constraints are not taken into account in the isoneutral diffusion and according to our interpretation this comes to consider uncertainties  that are not volume preserving.

In the case of  the Kraichnan model the density variation  is simplified as
\begin{equation}
\dif_t \rho + {\nab}\rho\,{\bm\cdot}\, \w \,\dif t  - \gamma \frac{1}{2}\sum_{i,j} \Delt \rho\, \dif t  = \nab \rho \,{\bm\cdot}\, \bsigma \,\dif\hat \B_t,
\end{equation}
and when the drift term does not depend on random argument  ({\it i.e.} if it corresponds to the flow expectation $\w=\Exp \dif\mbf X_t$ as in the case of a mean field dynamics) the density  expectation evolution is a classical intuitive advection diffusion equation
\begin{equation}
\upartial_t \bar{\rho} + {\nab}\bar{\rho}\,{\bm\cdot}\,\w     =  \tfrac{1}{2}\gamma \Delt \bar{\rho}.
\end{equation}

The mass conservation constraint and the stochastic version of the Reynolds theorem allows us now expressing  the linear momentum conservation  equations. 
\section{Conservation of momentum}
Newton's second law states that  the net force acting on the fluid is equal to the rate of change of the linear momentum. In our case, the flow evolution is described  through an Ito diffusion (\ref{Ito1}), where the drift component $\w(\xx,t)$ represents the unknown flow we wish to estimate whereas the noise term denotes the fluctuating part either caused  by physical processes that are neglected  or generated by numerical errors and  coarsening processes. Whatever the type of uncertainties considered on the fluid particles location, they introduce {\em de facto} an uncertainty on the forces. In order to take into account all the uncertainty sources within the momentum equations we consider a stochastic conservation principle of the form
\begin{equation}
\dif\!\!\int_{{\cal V}(t)} \!\!\!\!\!\!\!\!{\rho\bigl(\w(\xx,t)\dif t +\bsigma(\xx,t)\dif \hat\B_t\bigr)\dif \xx}=  \int_{{\cal V}(t)}\!\!\!\!\!\!\!\!\!\boldsymbol{F}(\xx,t)\dif \xx.
\end{equation}
In this equation the acceleration is highly irregular and has to be interpreted in the sense of distribution. For every $h\in C_0^\infty(\mathbb{R}+)$:
\begin{multline}
\label{eq-dist1} 
 \int h(t) \int_{{\cal V}(t)}\!\!\!\!\!\!\!\!\!\boldsymbol{F}(\xx,t) \dif \xx \,\dif t = - \int h'(t) \int_{{\cal V}(t)}\!\!\!\!\!\!\!\!\!\bsigma(\xx,t)\dif \hat\B_t\dif \xx\, \dif t+ \int h(t) \,\dif\!\! \int_{{\cal V}(t)}\!\!\!\!\!\!\!\!\!\rho\w(\xx,t)\dif \xx \,\dif t.
\end{multline}
Since both sides of this equation must have the same structure, the forces can be written as
\begin{multline}
\label{eq-dist2} 
\int h(t) \int_{{\cal V}(t)}\!\!\!\!\!\!\!\!\!\boldsymbol{F}(\xx,t) \dif \xx \,\dif t = - \int h'(t) \int_{{\cal V}(t)}\!\!\!\!\!\!\!\!\!\bsigma(\xx,t)\dif \hat\B_t\dif \xx \,\dif t \\[-0.3em]
+ \int h(t) \int_{{\cal V}(t)}\!\!\!\!\!\!\!\bigl(\boldsymbol{f}(\xx,t) \dif \xx \dif t + \boldsymbol{\theta}(\xx,t)\dif \hat\B_t\bigr)\dif \xx.
\end{multline}
The first terms of the right-hand sides of equations (\ref{eq-dist1}) and (\ref{eq-dist2}) are identical and cancel out. The second right-hand term of equation (\ref{eq-dist1}) corresponds to the derivative of the momentum associated to the resolved velocity component, whereas the second right-hand term of (\ref{eq-dist2}) provides us the structure of the forces under localization uncertainty. We now further develop those different terms. 

According to the previous section results, the transport equation applied to the linear momentum gives for each component of the velocity:
\begin{multline}
\dif\!\!\int_{{\cal V}(t)}\!\!\rho w_{i} \dif \xx =\int_{{\cal V}(t)}\!\!\biggl[\Bigl( \dif_t( \rho w_{i}) +  {\nab} {\bm\cdot}\,(\rho w_{i}\w)+\| \nab{\bm\cdot} \,\bsigma \|^2  \rho w_{i} - \sum_{j,k}\frac{1}{2}\frac{\upartial^2}{\upartial x_j\upartial x_k}(a^{jk}\rho w_{i}){|_{\nab{\bm\cdot}\, \bsigma=0}} \Bigr) \dif t  \biggr.\\[-0.6em]
\biggl.+\nab{\bm\cdot}\bigl(\rho w_{i}\bsigma \dif\hat \B_t\bigr)\biggr]\dif \xx.
\label{transport-momentum}
\end{multline}


As for the forces, we will consider that only  body forces and surface forces are involved (i.e., there is no external forces).
A direct extension of the deterministic case provides us the surface forces as
\begin{equation}
\label{sto-stress}
\mbs \Sigma= \int_{{\cal V}} - \nab (p\dif t +\dif\hat p_t) + \mu \Bigl(\Delt \mbs U + \frac{1}{3} \nab(\nab{\bm\cdot}\,\mbs U)\Bigr),\nonumber
\end{equation}
where $\mu$ is the dynamic viscosity, $p(\xx,t)$ denotes the deterministic contribution of the pressure and $\dif\hat p_t$ is a zero mean stochastic process describing the pressure fluctuations due to the random velocity component. 
The gravity force is deterministic and standard. 

Expressing the balance between those forces and  the acceleration (\ref{transport-momentum}), incorporating then the mass preservation principle (\ref{mass-conservation-1}), and finally equating, in the one hand, the slow temporal bounded variation terms and, in the other hand, the  Brownian terms, we get the expression of  the flow dynamics:
\begin{subequations}
\label{sto-NS}
\begin{align}
\Bigl(\frac{\upartial \w}{\upartial t} + \w{\nab}\transp \w \Bigr)\rho  -  \frac{1}{2} \sum_{i,j} a_{ij} \rho \frac{\upartial^2 \w}{\upartial x_i \upartial x_j} - &\sum_{i,j}\frac{\upartial (a_{ij} \rho)}{\upartial {x_j }}{|_{\nab{\bm\cdot}\, \bsigma=0}}  \frac{\upartial\w}{\upartial x_i}    \nonumber\\[-0.4em]
&\qquad= \rho \mbs g  - \nab p + \mu \Bigl(\Delt \w + \frac{1}{3} \nab(\nab{\bm\cdot}\,\w)\Bigr),\\
\nab \dif\hat p_t  =\!
 -   \rho\w\nab\transp\! \bsigma \dif \hat\B_t + \mu \Bigl(\Delt \Bigr.&\Bigl.\bigl(\bsigma \dif \hat \B_t\bigr) + \frac{1}{3} \nab\bigl(\nab{\bm\cdot}\,(\bsigma \dif\hat \B_t)\bigr)\Bigr),\\
\dif_t\rho \!+\! \Bigl({\nab} {\bm\cdot}\,(\rho \w )  - \frac{1}{2}\sum_{i,j} \frac{\upartial^2}{\upartial x_i \upartial x_j}(a_{ij}\rho)\Bigr.&\Bigl.{|_{\nab{\bm\cdot}\, \bsigma=0}} +\tfrac{1}{2}\| \nab{\bm\cdot}\, \bsigma \|^2  \rho\Bigr)\dif t ={\nab} {\bm\cdot}\bigl(\rho \bsigma \dif \hat\B_t\bigr).
\end{align}
\end{subequations}
This system provides a general form of the Navier Stokes equations under location uncertainty. Similarly to the Reynolds decomposition, the dynamics associated to the  drift component includes an additional stress term that depends on this resolved component and on the uncertainty diffusion tensor. In order to specify further the different terms involved in this general model, we will examine  simpler particular instances of this system.  In the following, we will confine ourselves to the case of an incompressible fluid of constant density. 
Considering as a first example the divergence free isotropic Kraichnan model defined in (\ref{Kxi})  and its associated transport equation,   we obtain straightforwardly the following Navier-Stokes equation:
\begin{subequations}
\label{NS-Kraichnan}
\be
 \Bigl(\frac{\upartial \w}{\upartial t} + \w{\nab}\transp \w  - \gamma \tfrac{1}{2} \Delt\w    \Bigr)\rho=
\rho \mbs g -  \nab p + \mu \Delt \w ,
\ee
\vskip -5mm
\be
\nab \dif\hat p_t  =- \rho (\w {\nab}\transp) \dif\boldsymbol{\xi}_t  + \mu \Delt \dif\boldsymbol{\xi}_t,
\ee
\vskip -5mm
\be
{\nab}{\bm\cdot} \,\w=0.
\ee
\end{subequations}
The first equation of this model corresponds to a traditional turbulent diffusion modeling relying on the Boussinesq assumption with a constant eddy viscosity coefficient.  Let us note that for this system the second equation related to the  random pressure term is not needed to compute the resolved drift $\w$. The uncertainty  is in that case specified  {\em a priori} as an homogeneous random field. Considering a more general divergence free  random component, we obtain
\bme
\label{NS-div-free}
\be
\se
\Bigl(\frac{\upartial \w}{\upartial t} + \w{\nab}\transp \w    \! -\! \frac{1}{2}\sum_{i,j}   \upartial_{x_i}\upartial_{x_j} (a^{ij} \w) \Bigr) \rho=
\rho \mbs g  \! -\! \nab p + \mu \Delt \w,
\ee
\vskip -5mm
\be
\se
\nab \dif\hat p_t  =
 - \rho (\w {\nab}\transp) \mbs \sigma \dif \hat\B_t  + \mu \Delt \mbs \sigma \dif \hat\B_t,
\ee
\vskip -5mm
\be
\te
\nab{\bm\cdot}\, \w=0,\qquad\qquad \nab{\bm\cdot}\,\bsigma=0,\qquad\qquad\nab{\bm\cdot}\,(\nab{\bm\cdot} \,\mbf a) = 0.
\ee
\eme
This model of Navier-Stokes equations includes now a  diffusion term that cannot be directly related to the traditional Boussinesq eddy viscosity formulation  anymore. Furthermore, the incompressibility condition on the variance tensor provides a non-local constraint on the subgrid model, which is lacking in usual eddy viscosity models \citep{Kraichnan87}. One can point out  that for divergence-free incompressible uncertainty models this term is  globally dissipative. As a matter of fact the energy associated to the subgrid term reads
\begin{align}
\int_\Omega \w\transp \sum_{i,j} \frac{\upartial^2}{\upartial x_i \upartial x_j}(a_{ij}\w) = & \int_\Omega \sum_i \frac{\upartial}{\upartial x_i}\biggl(\w\transp  \sum_j \Bigl(\frac{\upartial}{\upartial {x_j}}a_{i j} \w\Bigr)\biggr) \nonumber\\  & - \int_\Omega \sum_k \nab w_k\transp \mbs a \nab w_k  - \int_\Omega (\nab{\bm\cdot}\, \mbs a)\sum_k \nab w_k w_k \nonumber\\
=& - \int_\Omega \bigl\| \nabla \w \bigr\|^2_{\mbs a} - \int_\Omega (\nab {\bm\cdot}\, \mbs a) \nab \bigl(\tfrac{1}{2}\|\w\|^2\bigr).
\label{SGTKE}
\end{align}
The first term $\|\nabla \w \|^2_{\mbf a}= \sum_{ij} a_{ij} \upartial_{x_i}\w\,{\bm\cdot}\, \upartial_{x_j}\w$,  is always non-negative  as,  $\mbs a=\bsigma\bsigma^T$, is semi-positive definite. The second term associated to the incompressibility constraint, $\nab{\bm\cdot}\nab \mbf a =0$, cancels:
\begin{align}
\int_\Omega (\nab{\bm \cdot}\, \mbf a) \nab \bigl(\tfrac{1}{2}\|\w\|^2\bigr)\dif \xx &= \frac{1}{2}\int_\Omega \Bigl(\nab{\bm\cdot}\bigl( \nab{\bm\cdot}\,\mbf a\;\|\w\|^2\bigr)\Bigr) \dif \xx\\
&=\frac{1}{2}\int_{\upartial\Omega} \|\w\|^2 (\nab {\bm\cdot}\, \mbs a)\,{\bm\cdot} \, \mbf n \dif s\,=\,0\,.\nonumber
\end{align}
 The energy of the subgrid term reduces to
\begin{equation}
\int_\Omega \w\transp \sum_{i,j} \frac{\upartial^2}{\upartial x_i \upartial x_j}(a_{ij}\w) \dif \xx= - \int_\Omega \bigl\| \nab \w \bigr\|^2_{\mbf a}\dif \xx, 
\end{equation}
and is thus globally dissipative. Such model paves the way to set up subgrid models and stochastic expression of Navier-Stokes equations for different forms of the diffusion tensor.
It can be noted that this subgrid model provides the same dissipation as
\begin{equation}
\int_\Omega \w\transp \sum_{i,j} \upartial_{x_i} (a^{ij} \upartial_{x_j}\w)\dif \xx= - \int_\Omega \bigl\| \nab \w \bigr\|^2_{\mbf a}\dif \xx,
\end{equation}
which involves a diffusion operator ${\cal D} f=\sum_{i,j} \upartial_{x_i} (a^{ij} \upartial_{x_j} f)$ similar to diffusion  introduced in LES \citep{Karamanos00} through the concept of spectral vanishing viscosity operator  initially introduced for 1D problems \citep{tadmor89}. No clear consensus, nevertheless, exists in the 3D case on the form such an operator  should take \citep{Pasquetti06}.

\subsection{Kinetic energy of the complete flow}
From this expression of subgrid stress energy it is insightful to exhibit the total kinetic energy associated to whole flow. This expression will allow us to set an additional constraint on the energy of the random unresolved  process.  
For a divergence free turbulent component, the kinetic energy associated to the resolved drift component is first of all provided by
\begin{equation}
\frac{1}{2}\int_{\Omega}\upartial_t|\w|^2+\nu\int_{\Omega}|\nab\w|^2+\frac{1}{2}\sum_{i,j}\int_{\Omega}a_{ij}\upartial_{x_i}\w\,{\bm \cdot}\,\upartial_{x_j}\w= \big(\mbs f,\w\big).
\end{equation}
Here $\mbs f $ gathers the whole external forcing. From this expression and the energy of the uncertainty component, we get the mean kinetic energy of the global velocity field $\mbs U(\xx,t)  = \w(\xx,t)\dif t + \bsigma(\xx,t) \dif\hat \B_t$: 
\begin{subequations}
\begin{align}
\frac{1}{2\dif t}\Exp \int_\Omega &|U|^2 =  \frac{1}{2}\int_\Omega \int_0^t \upartial_t |\w|^2 \dif t +\frac{1}{2} \int_\Omega \trans( \mbs a) \dif \xx\\
 = &\frac{1}{2} \int_\Omega \trans (\mbs a)  \dif \xx - \int_0^t  \biggl\{\int_\Omega \nu|\nab\w|^2+\frac{1}{2}\sum_{i,j}\int_{\Omega}a_{ij}\upartial_{x_i}\w\,{\bm\cdot}\,\upartial_{x_j}\w -\big(\mbs f,\w \big) \biggr\}\dif t.
\end{align}
\end{subequations}
For inviscid flows without external forcing this energy should be conserved. Differentiating with respect to time we have thus: 
\begin{align}
\label{Evol-ener}
\upartial_t \int_\Omega \trans (\mbs a) \dif \xx -\sum_{i,j}\int_{\Omega}a_{ij}\upartial_{x_i}\w\,{\bm\cdot}\,\upartial_{x_j}\w =0.
\end{align}
This equation provides us a simple evolution model of the uncertainty component energy. It expresses a balance between the diffusive damping term that drains energy from the resolved component and the energy of the stochastic process that is back-scattered to the resolved smooth component. For the  Kraichnan uncertainty model, denoting $\gamma(t) = \int_\Omega \trans (\mbs a(t)) \dif \xx$ the uncertainty field energy, we get, the first order ordinary equation:
\begin{equation}
\upartial_t \gamma(t) = \gamma(t) \bigl\| \nab \w \bigr\|^2_2,
\end{equation}
which implies $\gamma(t) = \gamma(0)\exp \int_0^t\|\nab \mbs w \|^2_2 \dif t$. The uncertainty energy grows exponentially in time with respect to the velocity gradient norm. For decaying turbulence this quantity stays bounded. With an additional ellipticity condition on the quadratic variation tensor, it can be shown through classical arguments that the drift gradient norm is bounded by the initial condition and the external force (see appendix \ref{A}).  When the resolved velocity gradient increases the diffusion get stronger. The velocity gradient are in return smoothed accordingly and a larger scale representation for the drift flow  is obtained. For non isotropic diffusion tensors the smoothing is operated in an anisotropic way. Its intensity depends on an oriented velocity gradient norm. 

\subsection{Link to the Smagorinsky subgrid model}
The Smagorinsky subgrid model is widely used in large eddies simulation approaches. This eddy viscosity model, built from the Boussinesq assumption, imposes that the proper directions of the subgrid stress tensor are strictly aligned with those of the resolved rate of strain tensor.  The subgrid model derived from our uncertainty analysis does not rely on such assumption. Nevertheless,  a relation with the Smagorinsky model  term can be easily emphasized. As a matter of fact, setting the quadratic variation tensor to the form
\begin{equation}
\mbf a = C\|\mbs S\|\Id, 
\end{equation}
with $C$ a constant and where $\|\mbs S\|= \frac{1}{2}[\sum_{ij}(\upartial_{x_i} w^j + \upartial_{x_j} w^i)^2]^{1/2}$ denotes the Frobenius norm of the rate of strain tensor, we get an uncertainty stress tensor term that reads
\begin{subequations}
\begin{align}
\sum_{ij}\upartial_{x_i}\upartial_{x_j} (a^{ij}  w^k) &= \sum_{ij} \upartial_{x_i}  \upartial_{x_j}\bigl(\|\mbs S\|\delta^{ij}   w^k\bigr)\\
 &= 2\sum_j \upartial_{x_j} \|\mbs S\| \upartial_{x_j} w^k + \|\mbs S\|\Delt w^k + \Delt\|\mbs S\| w^k.
\end{align}
\end{subequations}
This term complemented by $2\sum_j \upartial_{x_j} \|\mbs S\|  \upartial_{x_k} w^j - \Delt\|\mbs S\| w^k$ provides the standard trace free Smagorinsky subgrid stress,  $\nab{\bm\cdot}\, {\bm \tau}$, where
\begin{equation}
{\bm \tau}= C\|\mbs S\| \mbs S.
\end{equation}
The complementary term may be rewritten as 
\begin{equation}
2\upartial_{x_k} \sum_j \upartial_{x_j} (\|\mbs S\|)   w^j - 2\sum_j \upartial_{x_j}\upartial_{ x_k} (\|\mbs S\|)   w^j- \Delt\|\mbs S\| w^k.
\end{equation}
We observe that the first term is a gradient term that can be compensated by a modified pressure as it is usually done considering deviatoric expression of the Reynolds stress tensor. However, unless assuming that the rate of strain magnitude is a very smooth function with flat iso-surfaces  there is no particular reason to cancel the two other terms. Let us note this condition allows  fulfilling the variance  incompressibility  constraint $\nab{\bm\cdot}\nab {\bm\cdot}\, \mbf a =0$. According to our interpretation, the Smagorinsky model constitutes thus an adapted uncertainty model only for smooth deformations with flat rate of strain tensor norm.

In the examples provided previously, the uncertainties have been fixed {\em a priori}. In such situations the second equation of system (\ref{NS-div-free}) is not required unless a realization of the oscillating random pressure field is needed (which is a direct possibility of our formulation). However, when the diffusion tensor is not specified  the system exhibits another degree of freedom and this tensor or at least the quadratic variation tensor must be estimated in order to fulfill the balance of Newton's second law. 
\subsection{Diffusion and quadratic tensors estimation}
The second equation of (\ref{NS-div-free}) provides a mean to proceed to the estimation of the diffusion tensor. Projecting this equation on the divergence-free space tensor, for a given increment $\mbs\epsilon=\yy-\xx$, we obtain a matrix screened Poisson equation with variable source functions:
\begin{equation}
\Delt\bsigma(\xx,\mbs\epsilon+\xx) =\frac{\rho}{\mu} {\cal P}\star (\w\nab\transp \bsigma)(\xx,\mbs\epsilon+\xx),
\end{equation}
where ${\cal P}$ indicates the divergence free projector, applied to the column vectors of $\w\nab\transp\bsigma$. A symmetric solution can be imposed adding and averaging the transposed problem on both sides: 
\begin{equation}
\Delt\bsigma(\xx,\mbs\epsilon+\xx) =\frac{\rho}{2\mu} [{\cal P}\star (\w\nab\transp \bsigma)+ (\bsigma \nab \w\transp )\star {\cal P}](\xx,\mbs\epsilon+\xx).
\end{equation}
Formally this equation should be solved for each increment value, which is obviously unrealistic in practice. Nevertheless, with the hypothesis of an homogeneous decorrelation function the resolution can be led only for the diagonal ($\mbs \epsilon=0$); the complete tensor $\bsigma(\xx,\yy)$ is then inferred  from those values and the decorrelation function.  Another solution consists to assume a rapid decorrelation and to solve the system for a narrow band around the diagonal.  An iterative solution can be built assuming an initial diffusion tensor, and solving these equations from the current values of $\w$ and $\bsigma$. Let us point out that the quadratic variation tensor, $\mbs a$, requires  finally an ultimate projection to stick to the last constraint of the system (\ref{NS-div-free}). A normalization and a  weighting  with respect to (\ref{Evol-ener}) enables  then to  guaranty the balance between the drained energy and the backscattered unresolved energy.

\section{Shallow water model}
We describe in this section an application of our  formalism to the derivation of a stochastic model of the shallow-water equations. 

The Shallow water equations constitute one of the simplest model that can be used to describe the evolution of mean horizontal components of atmospheric winds or oceanic streams.  This 2D model is valid for problems in which the vertical dynamics can be neglected compared to the horizontal effects and cannot be used when 3D effects become essential. It is derived from a 3D incompressible model by depth averaging assuming the depth of the vertical domain is "shallow" compared to the horizontal domain.  In order to establish the Shallow water approximation, the pressure  variable is assumed to follow quite correctly an hydrostatic equilibrium relation given by
\begin{equation}
\frac{\upartial p}{\upartial z} = -g \rho,
\label{eq-hydro}
\end{equation}
where the density is constant on a shallow layer of the fluid. 

In our case, the whole pressure  function is given as the summation of a deterministic pressure contribution, $p'$, and, $\dif \hat p$, a zero mean random pressure
\begin{equation}
p(x,y,z,t)= p'(x,y,z,t) + \dif \hat p(x,y,z,t).
\label{inc-pressure}
\end{equation}
 Assuming the external force is due to gravity and neglecting the friction forces, the hydrostatic balance for a deterministic system can be regarded as a boundary layer assumption in which the vertical acceleration of the fluid is suppose to be null. Considering a system with uncertainty under the same hypotheses, that is to say: no friction and a vertical acceleration compensated by the subgrid diffusion,    we obtain an identical hydrostatic relation (\ref{eq-hydro}). However, two different choices are possible. Either the hydrostatic balance can be assume to hold for the global pressure or only for the deterministic component. In the first case, in complement to an hydrostatic balance on the deterministic pressure component,  the random pressure component is necessarily such that
\begin{equation}
\upartial_z \dif \hat p_t =0.
\label{eq-hydro-turb-pressure}
\end{equation}
A strong hydrostatic assumption on the global pressure component hence requires that the random pressure term is constant along the vertical axis. In the second case, there is a supplementary degree of freedom and only a boundary condition on the random component will have to be imposed.

With the assumption that  the vertical shear on the horizontal  velocity fields is negligible,  which is reasonable for a fluid with shallow hypothesis,  (assuming also no rotation for sake of simplicity) and using (\ref{eq-hydro}) we obtain a 2D momentum equation that reads
\bme
\label{UNS-2}
\be
\se
\biggl(\frac{\upartial \w^h}{\upartial t} + \w^h{\nab^h}\transp \w^h  - \frac{1}{2}\sum_{i,j} \frac{\upartial^2}{\upartial x_i \upartial x_j}\bigl(a_{ij}\w^h\bigr)\biggr)\rho  =
  - \nab^h p',
\ee
\vskip -3mm
\be
\se
\nab^h \dif \hat p_t  =
 - \rho \bigl(\w^h {\nab^h}\transp\bigr) \bigl(\bsigma \dif \hat  \B_t\bigr)^h,
\ee
\be
\te
{\nab}{\bm \cdot} \,\w=0, \qquad\qquad
{\nab} {\bm \cdot} \,\bsigma =0,\qquad\qquad
{\nab}{\bm \cdot} {\nab} {\bm \cdot}\, \mbf a  =0.
\ee
\eme
Here the superscript, $h$, indicates an horizontal vector. 
In order to stick to the no vertical shear hypothesis the horizontal velocity components can be seen as averaged along the vertical axis between an upper free surface, $h_u$, of the fluid and the bottom topography, $h_b$. As for the turbulent component it is natural in the same way  to rely on a 3D Brownian uncertainty vector independent of the depth. Its quadratic variation consequently  does not depend on depth either. This choice ensure a strict respect of a null vertical shear and provides a diffusion tensor that is independent from depth. The 3D velocity will be thus defined as a velocity field of the form
\begin{equation}
\begin{pmatrix}
U(x,y,t) \\[0.8em] V(x,y,t) \\[0.8em] W(\xx,t)
\end{pmatrix}
=
\begin{pmatrix}
u(x,y,t)\dif t + \bigl(\bsigma(x,y,t) \dif\hat \B^h_t\bigr)_x\\[0.4em]v(x,y,t)\dif t + \bigl(\bsigma(x,y,t) \dif\hat \B^h_t\bigr)_y\\[0.4em] w(\xx,t)\dif t+ \bigl(\bsigma(x,y,t)\dif \hat \B^h_t\bigr)_z
\end{pmatrix}.
\end{equation}
Notice that the noise is a 3D white noise defined on the plane and that the vertical velocity component  depends on the height.
Considering the free surface, $h_u$, as a material surface in which no fluid crosses the surface, we have, from the stochastic transport principle,
\begin{equation}
\dif_t(h_u) + \biggl({\nab} h_u \w^h -\frac{1}{2} \sum_{(i,j)^h} \frac{\upartial^2}{\upartial x_i \upartial x_j}(a_{ij}h_u)\biggr) \dif t + \nab{\bm\cdot}\Bigl(h_u \bigl(\bsigma \dif\hat \B_t\bigr)^h\Bigr)= -w(h_u) - \bigl(\bsigma \dif\hat \B_t\bigr)_z,
\label{hu}
\end{equation}
where, $w(h_u)$, denotes the vertical velocity at point $(x,y)$ of the surface $h_u$ and $(i,j)^h$ indicates summation over horizontal indices. In the same way, for the stationary topographic depth we obtain
\begin{equation}
\biggl({\nab} h_b \w^h -\frac{1}{2} \sum_{(i,j)^h} \frac{\upartial^2}{\upartial x_i \upartial x_j}(a_{ij}h_b)\bigg) \dif t + \nab{\bm\cdot}\bigl(h_b \bsigma \dif\hat \B_t\bigr)  = -w(h_b) - \bigl(\bsigma \dif\hat \B_t\bigr)_z.
\label{hb}
\end{equation}

The integration of the hydrostatic relations (\ref{eq-hydro}) from a depth $z$ up to the free surface, gives for a constant density:
\begin{equation}
\label{eq-p'}
p'(x,y,z,t) -p'(x,y,h_u,t ) = g \rho(h_u-z).
\end{equation}
Here enters the two possible choices on the hydrostatic balance. For a  hydrostatic balance defined up to a noise, it is possible to impose an isobaric upper boundary condition to the global pressure: $p_u(t)=p(x,y,h_u,t )$. This condition  implies necessarily  $p'(x,y,h_u,t )= p_u(t)$, accompanied with $\dif p(x,y,h_u,t )=0$ as a boundary condition for the random term. This last constraint is inappropriate for a global hydrostatic balance as in this case the random component is depth free. In this case, it is necessary to impose a boundary condition only to the deterministic pressure function. Let us note that in this latter case, (\ref{eq-hydro-turb-pressure}) imposes to the uncertainty to be orthogonal to the depth free vertical component gradient  $\upartial_z \dif p_t= w\nab\transp \bsigma \dif B_t=0$.  This also introduces a locally noisy variable pressure on the upper surface which might be difficult to control. The former case, is less strict and allows a supplementary degree of freedom for the uncertainty. 

 In both cases we have nevertheless from (\ref{eq-p'}) that
\begin{equation}
p'(x,y,z,t)  = g \rho(h_u-z) + p_u,
\end{equation}
and 
\begin{equation} 
\nab^{h} p'(x,y,z,t) = g\rho \nab^{h} h_u(x,y,t), 
\end{equation}
where we have assumed that $p_u$ is horizontally uniform.  Thus the momentum equation of the horizontal velocity field becomes 
\begin{equation}
\biggl(\frac{\upartial \w^h}{\upartial t} + \w^h{\nab}\transp \w^h  - \frac{1}{2}\sum_{(i,j)^h} \frac{\upartial^2}{\upartial x_i \upartial x_j}\bigl(a_{ij}\w^h\bigr)\biggr)\rho =  - g\rho\nab h_u,
\end{equation}
and is independent of the vertical coordinate. Integrating now the velocity drift divergence free constraint  along the $z$ axis , we get
\begin{equation}
-\nab{\bm\cdot} \,\w^h (h_u-h_b)= w(x,y,h_u,t)-w(x,y,h_b,t). 
\label{dp-const }
\end{equation}
In the same way, the integration along depth of the uncertainty divergence free constraint provides the relations
\bme
\label{int-div1-div2}
\begin{align}
\nab^h{\bm\cdot}\, \bsigma^h =& 0, &
\nab^h{\bm\cdot}\,(\nab^h{\bm\cdot}\, \mbf a^h) =& 0.
\end{align}
\eme
 The horizontal component of the uncertainty vector is hence divergence free and the horizontal part of the quadratic variation tensor obeys a 2D uncertainty mass preservation constraint.  
Introducing (\ref{int-div1-div2}a,b)
in the difference between (\ref{hu}) and (\ref{hb}) gives,  for $h=h_u-h_b$,
\begin{equation}
\dif_t h+ \Bigl({\nab} {\bm\cdot}\bigl(h \w^h\bigr) -\frac{1}{2} \sum_{(i,j)^h} \upartial_{x_i} \upartial_{x_j}  (a_{ij} h) \Bigr)\dif t + \nab h\bigl(\bsigma \dif\hat \B_t\bigr)^h=0.
\end{equation}
The 2D  whole shallow water model with uncertainty is  hence finally given by
\bme
\label{SW}
\be
\se
\Bigl(\frac{\upartial \w^h}{\upartial t} + \w^h{\nab}\transp \w^h  - \frac{1}{2} \sum_{(i,j)^h}   \upartial_{x_i} \upartial_{x_j} \bigl(a_{ij} \w^h\bigr)\Bigr)\rho =   - g\rho\nab h_u,
\ee
\vskip -3mm
\be
\se
\dif_t h+ \Bigl(({\nab} {\bm \cdot}\,\bigl(h \w^h\bigr) -\frac{1}{2} \sum_{i,j} \upartial_{x_i} \upartial_{x_j} (a_{ij} h)\Bigr) \dif t + \nab h\bigl(\bsigma \dif\hat \B_t\bigr)^h=0, 
\ee
\vskip -3mm
\be
\se
\nab^h \dif\hat p  =
 - \rho \bigl(\w^h {\nab}\transp\bigr) \bigl(\bsigma \dif\hat  \B_t\bigr)^h,
\ee
\vskip -3mm
\be
\nab {\bm\cdot}\,\bsigma^h = 0 ,\qquad\qquad
\nab {\bm\cdot}\bigl(\nab {\bm\cdot} \,\mbf a^h\bigr) = 0.
\ee
\eme
The surface is here driven by a stochastic pdes that involves a multiplicative noise. The horizontal velocity fields is coupled to this stochastic dynamics through the upper surface gradient. This system is nevertheless unclosed. As a matter of fact, in this inviscid case, the second equation cannot be used anymore to defined the diffusion tensor as one can always find a random pressure for any given diffusion tensors. The diffusion tensor or its quadratic variation must hence be modeled. 

We can point out it is possible to simplify greatly this system if one seeks only a mean horizontal velocity  field. As a matter of fact, taking the surface conditional expectation, $\bar h_u$,    upon  a given initial condition for the surface provides us the following continuity equation for the shallow water model with uncertainty:
\begin{equation}
\label{SW-cont-a}
\frac{\upartial \bar h}{\upartial t} + {\nab} {\bm\cdot}\,\bigl(\bar h \w^h\bigr) -\frac{1}{2} \sum_{(i,j)^h} \upartial_{x_i} \upartial_{x_j} (a_{ij} \bar h) =0,
\end{equation}  
 This amounts then to solve the system
\begin{subequations}
\be
\Bigl(\frac{\upartial \w^h}{\upartial t} + \w^h{\nab}\transp \w^h  - \frac{1}{2}\sum_{(i,j)^h} \upartial_{x_i} \upartial_{x_j} \bigl(a_{ij}\w^h\bigr)\Bigr)\rho =   - g\rho\nab \bar h_u,
\ee
\vskip -3mm
\be
\frac{\upartial \bar h}{\upartial t} + {\nab} {\bm\cdot}\,\bigl(\bar h \w^h\bigr) -\frac{1}{2} \sum_{(i,j)^h} \upartial_{x_i} \upartial_{x_j}(a_{ij}\bar h) =0, 
\ee
\vskip -3mm
\be
\nab {\bm\cdot}\bigl(\nab {\bm\cdot} \mbf \,a^h\bigr) = 0.
\ee
\end{subequations}
The same system can also be  obtained specifying the uncertainties lie on iso-height surfaces since in that case the Brownian random terms cancel out. In both cases we get a deterministic formulation of the shallow water equations under uncertainty.  The first momentum and the free surface evolution include both a similar subgrid diffusion model accounting for the action of the uncertainty term associated to the  noisy fluid particle location.  Proceeding to an integration  along the depth direction  and writing the system in conservative form we finally get
\begin{subequations}
\be
\biggl(\frac{\upartial \bar h \w^h}{\upartial t} + \nab{\bm\cdot}\, \Bigl(\bar h\w^h\bigl(\w^h\bigr)\transp +g\tfrac{1}{2}\bar h^2\Id\Bigr)  -\sum_{(i,j)^h} \frac{1}{2} \upartial_{x_i}  \upartial_{x_j} \bigl(\bar h a_{ij}\w^h\bigr) -  a_{ij}\upartial_{x_j}\! \bar h \upartial_{x_i} \w\biggr)\rho \qquad\qquad\nonumber
\ee
\vskip -5mm
\be
\qquad\qquad\qquad\qquad \qquad\qquad =    - g\rho \bar h \nab \bar h_b,
\ee
\be
\frac{\upartial \bar h}{\upartial t} + {\nab} {\bm\cdot}\,\bigl(\bar h \w^h\bigr) -\frac{1}{2} \sum_{(i,j)^h} \upartial_{x_i} \upartial_{x_j} (a_{ij}\bar h)=0, 
\ee
\vskip -3mm
\be
\nab {\bm\cdot}\,(\nab {\bm\cdot}\, \mbf a^h) = 0.
\ee
\end{subequations}
  A stochastic system of the same  form could as well be obtained from the spde's (\ref{SW}a-e). 

Let us outline that the derivation applied here for the Shallow water system is quite general. This type of methodology could probably be extended to the different forms of geophysical flow models. Such developments constitute perspective works we would like to explore.  
In the following section, we briefly describe another example of the eventual benefits brought by the application of our formalism. This concerns the establishment of reduced order dynamical systems. 

\section{Application to low order dynamical system modeling}
 The constitution of low order dynamical models to describe the evolution of fluid flows arouses an intense research interest in several domains ranging from climatic study \citep{Majda99}, flow control \citep{Bergmann08, Noack-Book10} or  atmospheric sciences \citep{Selten95,Franzke05}. Reduced order model are usually defined from a Galerkin projection on a reduced basis specified from experimental measurements or numerical simulation data. One of the simplest way to define such empirical basis functions stems from the Karhunen-Loeve decomposition (referred as proper orthogonal decomposition - POD - in the fluid mechanics domain). In this decomposition the basis functions encodes the direction of higher variance and are solutions of eigenvalues problems associated either to the two points correlation tensor or to the temporal
correlation tensor (see for instance \citep{Holmes96}). This expansion is easy to implement, and leads through a Galerkin projection to a system of ordinary differential equations that enables representing with only few modes the evolution of complex flows. Nevertheless, the corresponding reduced system  reveals often unstable in practice and does not allow operating long term forecasts even if it fits perfectly to the data \citep{Artana-JCP12,Noack05}. To mitigate structural instabilities, artificial damping terms are usually introduced in the reduced model. One of the classical solution consists
in adding a constant viscosity acting in the same way on all the POD modes \citep{Rempfer94}. Approaches involving different viscosity values for the set of modes have been also considered \citep{Rempfer96} and spectral vanishing dissipative model  \citep{Karamanos00} have been introduced. Those damping terms added to the flow kinematic viscosity enables improving the system's numerical stability but require often a proper tuning  of the set of parameters involved.  Some of the methods proposed may even require  fixing {\em a priori} a great number of parameters.  The principle followed by all those techniques consists in artificially reintroducing a lost dissipative mechanism attached to the mode discarded by the severe truncation operated at the Galerkin projection stage. The inclusion of uncertainty in the flow dynamics brings naturally such a  mechanism, with the additional advantage that the parameters can be properly fixed from the data. In the following section we recall briefly the principles of the proper orthogonal decomposition, and we show how a Galerkin projection  of the Navier-Stokes equations under uncertainty onto a reduced set of empirical modes enables to get a dynamical system  of low order.

\subsection{POD basis}\label{podbasis}
POD decomposition has been widely used by different authors as a technique to obtain approximate descriptions
 of the large scale coherent structures in laminar and turbulent flows. Given an ensemble  $\bu (\xx,t_i)$ of velocity fields obtained experimentally at $M$ different discrete instants,
  POD provides a set of  $M$ mutually orthogonal basis functions, or \textit{modes}, $\bphi_i (\xx)$,
  which are optimal with respect to the  kinetic energy. 

Considering such a decomposition enables to write the velocity field as an average $\bar{\bu}$ with fluctuations captured by a finite set of modes:
  \begin{equation}
  \bu(x,t)=\overline{\bu} + \sum_{i = 1}^{M} b_{i}(t) \bphi_i(x),
\label{pod}
  \end{equation}
   where, $\overline{f}$, denotes a temporal average of function $f$. 
Seeking an optimal finite energy representation  subspace for ${u}({\xx},t_i)$ on the domain $\Omega$ and along the sampling time comes to find an ensemble of functions of  $L^2(\Omega)$  that maximize
 \begin{equation}
 \label{PB-POD}
 \overline{ \bigl|(\bu,\bpsi)\bigr|_2^2}, \qquad\mbox { subject to } \quad (\bpsi,\bpsi)=1,
 \end{equation}
The solution, $\bphi$, of this constrained optimization problem satisfies the following eigenvalue problem
\begin{equation}
\label{eqRfi}
 \int_\Omega K(\xx,\xx') \bphi_k (\xx)\dif \xx = \bphi_k(\xx) \lambda_k,
 \end{equation}
\noindent
 with the spatial autocorrelation -- or two points correlation --  tensor $$K(\xx,\xx')= \overline{(\bu(\xx,t) \bu(\xx',t))}=\frac{1}{M}\sum_{i=1}^{M}\bu(\xx,t_i) \bu(\xx',t_i).$$
 This tensor is linear, positive semi-definite and self-adjoint. As a consequence, it admits by the Hilbert-Schmidt theorem a spectral representation, which guaranties a solution to problem  (\ref{PB-POD}). The eigenvalues are real and positive and by the Mercer theorem, the autocorrelation can be represented as the (uniformly convergent) series:
 \begin{equation}
 K(\xx,\xx')= \sum_{i=0}^{+\infty} \lambda_i \bphi_k(\xx) \bphi\transp_k (\xx').
 \end{equation}
The spatial and temporal symmetry of the representation in terms respectively of the temporal $\{b_i,\,\, i=1,\ldots,M \}$ or spatial $\{\bphi_i, \,\,i=1,\ldots,M \}$ expansions  allows  interchanging the two points correlation tensor with the two times correlation tensor \citep{Sirovich87}. Such a procedure is numerically advantageous  when the temporal dimension is lower than the spatial dimension. This new eigenvalue problem, whose eigenvectors are the temporal coefficients, $b_k$,  is obtained by expressing in (\ref{eqRfi}) the spatial modes as a linear combination of the velocity fields
\[
\bphi_k(\xx) = \frac{1}{\lambda_k} \overline{b_k(t) \bu(\xx,t)} \,.
\]
The eigenvalues of this new problem are identical to those obtained for the spatial modes \citep{Holmes96}. 

\subsection{Galerkin projection}
A Galerkin projection enables to rewrite a partial differential equation (PDE) system as a system of
ordinary differential equation (ODE). According to this
procedure, the functions which define the original equation are projected on a finite
dimensions subspace of the phase space (in this case, the subspace generated by the first
$s$ modes).

A Reynolds decomposition of the Navier-Stokes equation under uncertainty can be formulated by separating the deterministic drift component into a  mean $\bar{\w}$ and a fluctuating $\w'$ parts: $\w'(\xx,t)=\w(\xx,t) - \bar{\w}(\xx)$. It describes the evolution of the fluctuating deterministic part of the velocity fields. This system reads
\bme
\label{Reynolds-STO}
\se
\begin{align}
 \upartial_t {\w'}+\w' \nab\transp \bar{\w} +\bar{\w}\nab\transp \w'+\w' \nab\transp \w'-\overline{\w'\nab\transp \w'}& +  \frac{\nab p'} {\rho}-\nu\Delt(\bar{\w}+\w')
\nonumber\\
&-\frac{1}{2}\sum_{i,j}\!\! \upartial_{x_i}\upartial_{x_j} \bigl(a_{ij}(\w' +\bar{\w})\bigr) =0,
\end{align}
\be
\nab \dif\hat p_t  =
 - \rho (\w {\nab}\transp) \mbs \sigma \dif \hat\B_t  + \mu \Delt \mbs \sigma \dif \hat\B_t,
\ee
\be
\te
\nab{\bm\cdot} \,\w=0,\qquad\qquad
\nab{\bm\cdot}\, \bsigma=0,\qquad\qquad
\nab{\bm\cdot}\,(\nab{\bm\cdot} \,\mbf a) = 0.
\ee
\eme
Let us assume now that the deterministic drift component, $\w$, lives  in a subspace of finite dimension spanned by a set of basis functions $\Phi\defin\{\bphi_i,\,\,i = 1,\ldots,m\}$ determined from a  sequence of observed motion fields $\{\bu(\cdot,t_j), \,\,j=1,\ldots,f\}$ much longuer than the number of empirical modes $(m<f)$. The highly random oscillating terms $\bsigma \dif\B_t$ are assumed to live in  the complement orthogonal space spanned by the basis $\Phi_{cs}\defin\{\bphi_i,\,\, i= m+1,\ldots,f\}$. Applying a Galerkin projection of system (\ref{Reynolds-STO}) onto the truncated basis, $\Phi$, we get
\begin{multline}
\biggl(
 \upartial_t {\w'}+\w' \nab\transp \bar{\w} +\bar{\w}\nab\transp \w'+\w' \nab\transp \w'- \overline{\w'\nab\transp \w'} +  \frac{\nab p'} {\rho}-\nu\Delt(\bar{\w}+\w') \biggr.\\\biggl.-\,\frac{1}{2}\sum_{i,j}\!\! \upartial_{x_i}\upartial_{x_j} \bigl(a_{ij}(\w' +\bar{\w})\bigr),\,\,\bphi_j \biggr)=0.
\label{GprojNS}
\end{multline} 
Rewriting (\ref{GprojNS}) in terms of the POD temporal coefficients, we obtain a quadratic system of ODEs. For every $j \leq m$ modes, the system reads
\begin{equation}
\label{ode}
\frac{\dif b_k}{\dif t} + i_k + \sum_{i=1}^m l_{ik} b_i + \sum_{i=1}^{s}\sum_{j= i}^{m} b_i c_{ijk} b_j =0\quad \forall k=1,\ldots, m\,,
\end{equation}
where
\bme
\se
\begin{align}
l_{i j} &=  \int_\Omega \bar{\w}\nab\transp \bphi_i \, {\bm\cdot}\,\bphi_j \dif \xx+ \int_\Omega  \bphi_i \nab\transp \bar{\w}\, {\bm\cdot}\,
\bphi_j \dif \xx -\frac{1}{2}\int_\Omega \sum_{k,\ell} \upartial_{x_k}\upartial_{x_\ell} (a_{k\ell} \bphi_i) \, {\bm\cdot}\,\bphi_j \dif \xx \qquad\qquad \nonumber\\[-0.5em]
&\qquad\qquad\qquad\qquad\qquad\qquad\qquad\qquad\qquad\qquad\qquad\qquad
- \int_\Omega\nu\Delt \bphi_i\, {\bm\cdot}\,\bphi_j \dif \xx,
\label{LIN}\\
\!\!c_{ijk}&= \!\int_\Omega \bphi_j \nab\transp \bphi_i \, {\bm\cdot}\,\bphi_k \dif \xx,
\label{ADV}\\
i_k &= \int_\Omega \nab p' \, {\bm\cdot}\,\bphi_k \dif \xx - \nu \int_\Omega   \Delt \bar{\w} \, {\bm\cdot}\,\bphi_k \dif \xx - \frac{1}{2}\int_\Omega \sum_{i,j}\upartial_{x_i} \upartial_{x_j}(a_{ij} \bar{\w})\, {\bm\cdot}\,\bphi_k \dif \xx  \nonumber\\[-0.5em]
&\qquad\qquad\qquad\qquad\qquad\qquad\qquad\qquad\qquad\qquad\qquad
- \sum_{j=1}^m \lambda_j  \int_\Omega \bphi_j \nab\transp \bphi_j\, {\bm\cdot}\, \bphi_k \dif \xx.
\label{CNT}
\end{align}
\eme

All those expressions capture different physical phenomenon. The first linear term  (\ref{LIN}) describes the interaction between the mean flow and the fluctuating field. It also includes viscous effects of the resolved modes and dissipation caused by the uncertainty component. Nonlinear advection effects are reported by (\ref{ADV}). The independent term (\ref{CNT}) takes into account mean flow dissipation due both to friction and to the action of the random uncertainty, convective contribution of the modes and the pressure field influence. Boundary conditions and symmetry make the pressure term vanish in the particular case of wake flows. As a matter of fact, each mode  satisfies the continuity equation, to give
\[ \int_\Omega \nab p'\, {\bm\cdot}\, \bphi_k \dif \xx= \oint_C p'  \bphi_k \dif s,\]
where $C$ is the boundary of domain $\Omega$. Works of \citep{Deane91} and \citep{Noack03} demonstrated that for wake flow configurations, the latter expression is negligible compared to the other terms. Direct calculation of each term of the system (\ref{ode}) can be avoided by relying on polynomial identification or optimal control techniques \citep{Artana-JCP12}. The later provide in addition a way to estimate an adapted initial condition of the system and to consider eventually an error term on the dynamics. Least squares identification or variational assimilation techniques \citep{Artana-JCP12}, hides  the necessity of an additional dissipation mechanisms as it is assumed to be carried by the data. However, even in that case it is useful to formulate in an explicit way the precise form of the uncertainty involved. 

The simplest procedure to specify the quadratic variation  tensor, $\mbs a$, associated to the uncertainty terms,  consists in identifying  it to the measurements variance:
\[
\boldsymbol a (\xx)= \frac{1}{T} \sum_{i=1}^{f}\biggl(\bu(\xx,t_i) - \bar{\bu}- \sum_{j=1}^{m} b_j(t_i)\bphi_j(\xx)\biggr)\biggl(\bu(\xx,t_i)-\bar{\bu}- \sum_{j=1}^{m} b_j(t_i)\bphi_j(\xx)\biggr){\biggl.\!\biggr.}\transp.
\]
Here it is assumed that the measurement variance tensor respects the incompressibility constraint $\nab{\bm\cdot}\nab{\bm\cdot}\, \mbs{a} = 0$. A strict respect of this constraint would require an additional projection defined through a constrained least squares fitting. However it can be noticed, that the simple scheme above provides a very intuitive scheme as the diffusion process of the large scale flow component is proportional to the residuals variance tensor. Locally the greater the discrepancy between the reduced model and the data the stronger the diffusion of the velocity components.

This procedure provides immediately an expression of the covariance tensor associated to the uncertainty in terms of the complement space basis functions  
 \begin{equation}
 \Exp \bigl(\bsigma(\xx) \dif\B_t (\bsigma(\xx') \dif\B_t)\transp\bigr)= \sum_{k=m+1}^{f} \lambda_k \bphi_k(\xx) \bphi\transp_k (\xx').
 \end{equation}
The uncertainty term can then be classically written in a spectral form as
\begin{equation}
\bsigma(\xx)\dif\B_t = \sum^{f}_{j=m+1} \lambda^{1/2}_j \bphi_j \dif\beta_j(t),
\end{equation}
where $\beta_j(t)$ denotes a set of uncorrelated scalar standard Brownian motions. This representation supplies thus a simple scheme for the uncertainty sampling and hence allows random drawings of the whole flow field:
\[
\dif\XX_t= \sum_{i=1}^{m} b_i(t)\bphi_i \dif t + \sum^{f}_{j=m+1} \lambda^{1/2}_j \bphi_j \dif\beta_j(t).
\] 
Such a formulation should enable to set up stochastic filtering procedures to estimate state variables of interest from partial observation like satellite images \citep{Beyou-Tellus13} in which the  reduced  order modeling would have the great advantage to offer the possibility of an  immediate significant augmentation of the ensemble members.

%
%
%

\section{Conclusion}
In this paper we have proposed a decomposition that allows us modeling the action on a resolved component of the flow dynamics of uncertainties related to the fluid particles displacement. The random uncertainty field encodes numerical artifacts or unresolved physical processes that have been neglected in the momentum balance. They are defined through diffusion tensors that have to be estimated or specified. In the former case, the estimation can be led from a set of vectorial Poisson equations and two divergence free constraints. In the latter case the system reduces to a Navier-Stokes equation with a subgrid modeling defined from an anisotropic diffusion that involves only the variance of the random term together with a global constraint on this variance. This formulation gives a way to define appropriate  subgrid diffusion schemes  from the flow uncertainties specification and might be useful to build large scale stochastic climatic or geophysical models.  It brings also a new point of view to some usual subgrid modeling and introduces additional constraints on those models. In this study such a  derivation has been investigated only for a shallow water model and relation of the subgrid term has been explained for the Smagorinsky model and for the Gent-McWilliams isopycnal diffusion. In the future we aim at exploring further the derivation of large scale geophysical models through such a methodology.  This includes for instance the rotating Boussinesq equations, or  quasi-geostrophic models. It would be also of particular interest to see if such a methodology enables to recover more realistic subgrid models that takes into account inhomogeneous flow over topography for the barotropic vorticity and baroclinic QG equations.   
Another perspective of work, will consist to investigate assimilation strategies between large scale models under uncertainty and fine resolution image data. The idea will be here to exploit the definition of the subgrid tensor in terms of statistical variances of the small scale velocities.

\appendix
\section{Global energy estimate}
\label{A}
 To get a global energy estimates, we will assume the field $\w$ is {smooth enough} to be taken as a test function, $\w\in H^1_0(\Omega)^d$ of a divergence free Sobolev space with null condition on the boundary for example. Then, taking $\w$ as test function in (\ref{NS-div-free}) we get the kinetic energy as
\begin{equation}
\frac{1}{2}\int_{\Omega}\upartial_t|\w|^2+\nu\int_{\Omega}|\nab\w|^2+\frac{1}{2}\sum_{i,j}\int_{\Omega}a_{ij}\upartial_{x_i}\w\,{\bm\cdot}\,\upartial_{x_j}\w= \langle \f,\w \rangle.
\end{equation}
An ellipticity condition on $a_{ij}$, 
\begin{equation}\label{ContrainteA1}
\sum_{i,j}a_{ij}\xi_i\xi_j\ge\alpha|\xi|^2 \qquad\forall~\xi\in\reel^d, \quad\mbox{with}\; \alpha>0 \,,
\end{equation}
leads to
\begin{equation}
\frac{1}{2}\int_{\Omega}\upartial_t|\w|^2+\tilde\nu\int_{\Omega}|\nab\w|^2\le \int_{\Omega}|\f||\w|,
\end{equation}
with
\begin{equation*}
\tilde\nu=\nu+\tfrac{1}{2}\alpha.
\end{equation*}
Since $\w\in H^1_0(\Omega)^d$, using Poincar\'e inequality, the term $\int_{\Omega}|\f||\w|$ can be split as follows:
 \begin{equation*}
\int_{\Omega}|\f||\w|\le\bigl\|\f\bigr\|_{H^{-1}(\Omega)^d}\bigl\|\w\bigr\|_{L^2(\Omega)^d}\le C(\Omega)\bigl\|\f\bigr\|_{H^{-1}(\Omega)^d}\bigl\|\nab\w\bigr\|_{L^2(\Omega)^d},
\end{equation*}
where $H^{-1}(\Omega)^d$ denotes the dual of $H^1_0(\Omega)^d$:
\begin{equation*}
\bigl\|\f\bigr\|_{H^{-1}(\Omega)^d}:=\sup_{\mbs v\in H^1_0(\Omega)^d}\frac{\langle\f,\mbs v\rangle}{\|\mbs v\|_{H^1_0(\Omega)^d}},  \quad \mbs v\ne0.
\end{equation*}
Young's inequality gives
\begin{equation*}
C(\Omega)\bigl\|\f\bigr\|_{H^{-1}(\Omega)^d}\bigl\|\nab\w\bigr\|_{L^2(\Omega)^d}\le \frac{\tilde\nu}{2}\bigl\|\nab\w\bigr\|^2_{L^2(\Omega)^d}+\frac{C(\Omega)^2}{2\tilde\nu}\bigl\|\f\bigr\|^2_{H^{-1}(\Omega)^d}.
\end{equation*}
Finally we get
\begin{equation*}
\int_{\Omega}\upartial_t|\w|^2+\tilde\nu\int_{\Omega}|\nab\w|^2\le\frac{C(\Omega)^2}{\tilde\nu}\bigl\|f\bigr\|^2_{H^{-1}(\Omega)},
\end{equation*}
and integration from $0$ to $T$ gives
\begin{equation*}
\bigl\|\w\bigr\|^2_{L^2(\Omega)^d}+\tilde\nu\int_0^T\bigl\|\nab\w\bigr\|^2_{L^2(\Omega)^{d\times d}}\le\bigl\|\w^{0}\bigr\|^2_{L^2(\Omega)^d}+\frac{C(\Omega)^2}{\tilde\nu}\int_0^T\bigl\|f\bigr\|^2_{H^{-1}(\Omega)},
\end{equation*}
thus
\begin{equation*}
\bigl\|\w\bigr\|^2_{L^2(\Omega)^d}\le\bigl\|\w^{0}\bigr\|^2_{L^2(\Omega)^d}+\frac{C(\Omega)^2}{\tilde\nu}\bigl\|f\bigr\|^2_{L^2(0,T;L^{2}(\Omega))},
\end{equation*}
and
\begin{equation*}
\tilde\nu\int_0^T\bigl\|\nab\w\bigr\|^2_{L^2(\Omega)^{d\times d}}\le\bigl\|\w^{0}\bigr\|^2_{L^2(\Omega)^d}+\frac{C(\Omega)^2}{\tilde\nu}\bigl\|f\bigr\|^2_{L^2(0,T;L^{2}(\Omega))}.
\end{equation*}
The kinetic energy and the gradient velocity norm are bounded by the external force $\f$ and the energy of initial condition $\w^{0}$.
\bibliographystyle{gGAF}

\begin{thebibliography}{48}
\providecommand{\natexlab}[1]{#1}

\bibitem[\protect\citeauthoryear{Artana
  {\itshape{et~al.}}}{2012}]{Artana-JCP12}
Artana, G., Cammilleri, A., Carlier, J. and M\'emin, E., Strong and weak
  constraint variational assimilations for reduced order fluid flow modeling.
  {\itshape J. Comp. Phys.} 2012, \textbf{231}, 3264--3288.

\bibitem[\protect\citeauthoryear{Bensoussan and
  Temam}{1973}]{Bensoussan-Temam-73}
Bensoussan, A. and Temam, R., Equations stochastique du type {N}avier-{S}tokes.
  {\itshape J. Funct. Anal.} 1973, \textbf{13}, 195--222.

\bibitem[\protect\citeauthoryear{Bergmann and Cordier}{2008}]{Bergmann08}
Bergmann, M. and Cordier, L., Optimal control of the cylinder wake in the
  laminar regime by trust-region methods and {POD} reduced-order models.
  {\itshape J. Comp. Phys.} 2008, \textbf{227}, 7813--7840.

\bibitem[\protect\citeauthoryear{Beyou
  {\itshape{et~al.}}}{2013}]{Beyou-Tellus13}
Beyou, S., Cuzol, A., Gorthi, S. and M\'emin, E., Weighted ensemble transform
  Kalman filter for image asssimilation. {\itshape Tellus A} 2013, \textbf{65}, 1--17.

\bibitem[\protect\citeauthoryear{Boussinesq}{1877}]{Boussinesq77}
Boussinesq, J., Essai sur la th\'eorie des eaux courantes. {\itshape M\'emoires
  pr\'esent\'es par divers savants \`a l'Acad\'emie des Sciences} 1877, \textbf{23} (1),
  1--680.

\bibitem[\protect\citeauthoryear{Brze\'{z}niak
  {\itshape{et~al.}}}{1991}]{Brzezniak91}
Brze\'{z}niak, Z., Capi\'{n}ski, M. and Flandoli, F., Stochastic partial
  differential equations and turbulence. {\itshape Math. Models Mathods Appl.
  Sci.} 1991, \textbf{1}, 41--59.

\bibitem[\protect\citeauthoryear{Constantin and Iyer}{2008}]{Constantin08}
Constantin, P. and Iyer, G., A stochastic {L}agrangian representation of the
  three-dimensionnal incompressible {N}avier-{S}tokes equations. {\itshape
  Comm. Pure Appl. Math.} 2008, \textbf{61}, 330--345.

\bibitem[\protect\citeauthoryear{Constantin and Iyer}{2011}]{Constantin11}
Constantin, P. and Iyer, G., A stochastic {L}agrangian approach to the of the
  {N}avier-{S}tokes equations in domains with boundary. {\itshape Ann. Appl. Probab.} 2011, \textbf{21}, 1466--1492.

\bibitem[\protect\citeauthoryear{Deane {\itshape{et~al.}}}{1991}]{Deane91}
Deane, A., Kevrekidis, I., Karniadakis, G. and Orszag, S., Low-dimensional
  models for complex geometry flows: Application to grooved channels and
  circular cylinders. {\itshape Phys. Fluids} 1991, \textbf{3}, 2337--2354.

\bibitem[\protect\citeauthoryear{Evensen}{2006}]{Evensen06}
Evensen, G., {\itshape Data assimilation: The ensemble {K}alman filter},  2006
  (Springer-Verlag, New-york).

\bibitem[\protect\citeauthoryear{Farge {\itshape{et~al.}}}{2001}]{Farge-PRL01}
Farge, M., Pellegrino, G. and Schneider, K., Coherent vortex extraction in 3D
  turbulent flows using orthogonal wavelets. {\itshape Phys. Rev. Lett.} 2001, \textbf{87(5)}, 054501.

\bibitem[\protect\citeauthoryear{Farge {\itshape{et~al.}}}{1999}]{Farge99}
Farge, M., Schneider, K. and Kevlahan, N., Non-Gaussianity and Coherent Vortex
  Simulation for two-dimensional turbulence using an adaptative orthogonal
  wavelet basis. {\itshape Phys. Fluids} 1999, \textbf{11}, 2187--2201.

\bibitem[\protect\citeauthoryear{Flandoli}{2008}]{Flandoli-08}
Flandoli, F., Vol.  1942 of {\itshape Lecture Notes in Math}, Lecture Notes in
  Math, An intoduction to 3D stochastic {N}avier {S}tokes. In {\itshape SPDE in
  hydrodynamics}, pp. 51--150, 2008 (Springer Verlag: Berlin).

\bibitem[\protect\citeauthoryear{Franzke and Majda}{2005}]{Franzke05}
Franzke, C. and Majda, A., Low-Order Stochastic Mode Reduction for a Prototype
  Atmospheric GCM. {\itshape J. Atmos. Sci.} 2005, \textbf{63}, 457--479.

\bibitem[\protect\citeauthoryear{Frederiksen}{1999}]{Frederiksen99}
Frederiksen, J., Subgrid-scale parameterizations of eddy-topographic force,
  eddy viscosity, and stochastic backscatter for flow over topography.
  {\itshape J. Atmos. Sci.} 1999, \textbf{56}, 1481--1494.

\bibitem[\protect\citeauthoryear{Frederiksen}{2012}]{Frederiksen12}
Frederiksen, J., Statistical Dynamical Closures and Subgrid Modeling for
  Inhomogeneous {QG} and {3D} Turbulence. {\itshape Entropy} 2012,
  \textbf{14}, 32--57.

\bibitem[\protect\citeauthoryear{Frederiksen
  {\itshape{et~al.}}}{2013}]{Frederiksen13}
Frederiksen, J., O'Kane, T. and Zidikheri, M., Subgrid modelling for
  geophysical flows. {\itshape Phil. Trans. R. Soc.} A, 2013, \textbf{371(1982)}, 20120166.

\bibitem[\protect\citeauthoryear{Gawedzky and Kupiainen}{1995}]{Gawedzki95}
Gawedzky, K. and Kupiainen, A., Anomalous scaling of the passive scalar.
  {\itshape Phys. Rev. Lett.} 1995, \textbf{75}, 3834--3837.

\bibitem[\protect\citeauthoryear{Gent and McWilliams}{1990}]{Gent90}
Gent, P. and McWilliams, J., Isopycnal mixing in ocean circulation models.
  {\itshape J. Phys. Oceanogr.} 1990, \textbf{20}, 150--155.



\bibitem[\protect\citeauthoryear{Gordon {\itshape{et~al.}}}{2001}]{Doucet01}
Gordon, N., Doucet, A. and Freitas, J.D., {\itshape Sequential {M}onte {C}arlo
  methods in practice},  2001 (Springer-Verlag).

\bibitem[\protect\citeauthoryear{Holmes {\itshape{et~al.}}}{1996}]{Holmes96}
Holmes, P., Lumley, J. and Berkooz, G., {\itshape Turbulence, coherence
  structures, dynamical systems and symetry},  1996 (Cambridge university
  press).


\bibitem[\protect\citeauthoryear{Karamanos and
  Karniadakis}{2000}]{Karamanos00}
Karamanos, G, and Karniadakis, G., A spectral vanishing viscosity method for
  large-eddy simulations. {\itshape J. Comp. Phys.} 2000,
  \textbf{163}, 22--50.

\bibitem[\protect\citeauthoryear{Kraichnan}{1959}]{Kraichnan59}
Kraichnan, R., The structure of isotropic turbulence at very high Reynolds
  numbers. {\itshape J. Fluid Mech.} 1959, \textbf{5}, 477--543.

\bibitem[\protect\citeauthoryear{Kraichnan}{1968}]{Kraichnan68}
Kraichnan, R., Small-scale structure of a randomly advected passive scalar.
  {\itshape Phys. Rev. Lett.} 1968, \textbf{11}, 945--963.

\bibitem[\protect\citeauthoryear{Kraichnan}{1970}]{Kraichnan70}
Kraichnan, R., Convergents to turbulence functions. {\itshape J. Fluid
  Mech.} 1970, \textbf{41}, 189--217.

\bibitem[\protect\citeauthoryear{Kraichnan}{1987}]{Kraichnan87}
Kraichnan, R., Eddy viscosity and diffusivity: exact formulas and
  approximations. {\itshape Complex Systems} 1987, \textbf{1}, 805--820.

\bibitem[\protect\citeauthoryear{Kunita}{1990}]{Kunita}
Kunita, H., {\itshape Stochastic flows and stochastic differential equations},
  1990 (Cambridge University Press).

\bibitem[\protect\citeauthoryear{Leith}{1971}]{Leith71}
Leith, C., Atmospheric predictability and two-dimensional turbulence. {\itshape
  J. Atmos. Sci.} 1971, \textbf{28}, 145--161.

\bibitem[\protect\citeauthoryear{Lesieur and M\'etais}{1996}]{Lesieur96}
Lesieur, M. and M\'etais, O., New trends in large-eddies simulation of
  turbulence. {\itshape Annu. Rev. Fluid. Mech.} 1996, \textbf{28}, 45--82.

\bibitem[\protect\citeauthoryear{Lilly}{1966}]{Lilly66}
Lilly, D., On the Application of the Eddy Viscosity Concept in the Inertial
  Subrange of Turbulence.  1966, Technical report 123, NCAR.

\bibitem[\protect\citeauthoryear{Majda and Kramer}{1999}]{Majda-Kramer}
Majda, A. and Kramer, P., Simplified models for turbulent diffusion:Theory,
  numerical modelling, and physical phenomena. {\itshape Physics report}, 1999,
  \textbf{314}, 237--574.

\bibitem[\protect\citeauthoryear{Majda {\itshape{et~al.}}}{1999}]{Majda99}
Majda, A., Timofeyev, I. and Vanden Eijnden, E., Models for stochastic climate
  prediction. {\itshape PNAS}, 1999, \textbf{96(26)}, 14687--14691.

\bibitem[\protect\citeauthoryear{Meneveau and Katz}{2000}]{Meneveau00}
Meneveau, C. and Katz, J., Scale-invariance and turbulence models for
  large-eddy simulation. {\itshape Annu. Rev. Fluid. Mech.} 2000, \textbf{32},
  1--32.

\bibitem[\protect\citeauthoryear{Mikulevicius and
  Rozovskii}{2004}]{Mikulevicius04}
Mikulevicius, R. and Rozovskii, B., Stochastic {N}avier-{S}tokes equations for
  turbulent flows. {\itshape SIAM J. Math. Anal.} 2004, \textbf{35},
  1250--1310.

\bibitem[\protect\citeauthoryear{Noack {\itshape{et~al.}}}{2003}]{Noack03}
Noack, B., Afanasiev, K., Morzynski, M., Tadmor, G. and Thiele, F., A hierarchy
  of low-dimensional models for the transient and post-transient cylinder wake.
  {\itshape J. Fluid Mech.} 2003, \textbf{497}, 335--363.

\bibitem[\protect\citeauthoryear{Noack {\itshape{et~al.}}}{2010}]{Noack-Book10}
Noack, B., Morzynski, M. and Tadmor, G. (Eds) {\itshape Reduced-Order Modelling
  for Flow Control},  Vol. 528, CISM Courses and Lectures 2010
  (Springer-Verlag).

\bibitem[\protect\citeauthoryear{Noack {\itshape{et~al.}}}{2005}]{Noack05}
Noack, B., Papas, P. and Monkevitz, P., The need for a pressure-term
  representation in empirical {G}alerkin models of incompressible shear flows.
  {\itshape J. Fluid Mech.} 2005, \textbf{523}, 339--365.

\bibitem[\protect\citeauthoryear{Palmer and Williams}{2008}]{Palmer08}
Palmer, T. and Williams, P., Theme Issue 'Stochastic physics and climate
  modelling'. {\itshape Phil. Trans. R. Soc.} A  2008, \textbf{366}.

\bibitem[\protect\citeauthoryear{Pasquetti}{2006}]{Pasquetti06}
Pasquetti, R., Spectral vanishing viscosity method for large-eddy simulation of
  turbulent flows. {\itshape J. Sci. Comp.} 2006, \textbf{27}, 365--375.

\bibitem[\protect\citeauthoryear{Prato and Zabczyk}{1992}]{DaPrato}
Prato, G.D. and Zabczyk, J., {\itshape Stochastic equations in infinite
  dimensions},  1992 (Cambridge University Press).

\bibitem[\protect\citeauthoryear{Rempfer}{1996}]{Rempfer96}
Rempfer, D., Investigation of of boundary layer transition via {G}alerkin
  projection on empirirical eigenfunctions. {\itshape Phys. Fluids} 1996,
  \textbf{8}, 175--188.

\bibitem[\protect\citeauthoryear{Rempfer and Fasel}{1994}]{Rempfer94}
Rempfer, D. and Fasel, H., Evolution of three-dimensional coherent structures
  in a flat-plate boundary layer. {\itshape J. Fluid Mech.} 1994,
  \textbf{260}, 351--375.

\bibitem[\protect\citeauthoryear{Sagaut}{2005}]{Sagaut04}
Sagaut, P., {\itshape Large-eddy simulation for incompressible flow - An
  introduction, third edition},  2005 (Springer-Verlag, Scientic Computation
  series).

\bibitem[\protect\citeauthoryear{Selten}{1995}]{Selten95}
Selten, F.M., An efficient description of the dynamics of barotropic flow.
  {\itshape J. Atmos. Sci.} 1995, \textbf{52}, 915--936.

\bibitem[\protect\citeauthoryear{Sirovich}{1987}]{Sirovich87}
Sirovich, L., Turbulence and the dynamics of coherent structures. {\itshape
  Quart. Appl. Math.} 1987, \textbf{45}, 561--590.

\bibitem[\protect\citeauthoryear{Slingo and Palmer}{2011}]{Slingo11}
Slingo, J. and Palmer, T., Uncertainty in weather and climate prediction.
  {\itshape Phil. Trans. R. Soc.} A 2011, \textbf{369}, 4751--4767.

\bibitem[\protect\citeauthoryear{Smagorinsky}{1963}]{Smagorinsky63}
Smagorinsky, J., General circulation experiments with the primitive equation:
  I. The basic experiment. {\itshape Mon. Weather Rev.} 1963,
  \textbf{91}, 99--165.

\bibitem[\protect\citeauthoryear{Tadmor}{1989}]{tadmor89}
Tadmor, E., Convergence of spectral methods for nonlinear conservation laws.
  {\itshape SIAM J. Numer. Anal.} 1989, \textbf{26}, 30--44.

\end{thebibliography}


\end{document}